\documentclass[a4paper,11pt]{article}
\pdfoutput=1 

\usepackage{jcappub} 

\usepackage[T1]{fontenc} 
\usepackage{graphicx}
\usepackage{epsfig}
\usepackage{rotate}
\usepackage{amsmath}
\usepackage{amssymb}
\usepackage{amsfonts}
\usepackage{bm}
\usepackage{tablefootnote}
\usepackage{enumerate}
\usepackage{afterpage}
\usepackage{xcolor}
\usepackage{natbib, hyperref}
\usepackage{multirow}
\UseRawInputEncoding

\newcommand{\ud}{\mathrm{d}}

\newcommand{\cH}{\mathcal{H}}
\newcommand{\fnl}{f_{\rm NL}}
\newcommand{\sfnl}{\sigma(f_{\rm NL})}

\def\be{\begin{equation}}
\def\ee{\end{equation}}
\def\bea{\begin{eqnarray}}
\def\eea{\end{eqnarray}}

\newcommand{\blu}[1]{{\color{blue}{#1}}}

\title{Constraining primordial non-Gaussianity by combining photometric galaxy and 21cm intensity mapping surveys}

\author{Mponeng Kopana$^1$, Sheean Jolicoeur$^{2}$, Roy Maartens$^{1,3,4}$}
\affiliation{$^{1}$Department of Physics \& Astronomy, University of the Western Cape, Cape Town 7535, South Africa\\
$^{2}$Department of Physics, Stellenbosch University, Matieland 7602, South Africa \\
$^{3}$Institute of Cosmology \& Gravitation, University of Portsmouth, Portsmouth PO1 3FX, United Kingdom\\
$^4$National Institute for Theoretical \& Computational Sciences (NITheCS), Cape Town 7535, South Africa }

\abstract{
The fluctuations produced during cosmic inflation may exhibit non-Gaussian characteristics that are imprinted in the large-scale structure of the Universe. This non-Gaussian imprint is an ultra-large scale signal that can be detected using the power spectrum. We focus on the local-type non-Gaussianity $f_{\rm NL}$ and employ a multi-tracer analysis that combines different probes in order to mitigate cosmic variance and maximize the non-Gaussian signal. In our previous paper, we showed that combining spectroscopic galaxy surveys with 21 cm intensity mapping surveys in interferometer mode could lead to a $\sim$ 20-30\% improvement in the precision on this non-Gaussian signal. Here we combine the same 21 cm experiments, including also single-dish surveys, with photometric galaxy surveys. The 21 cm single-dish surveys are based on MeerKAT and SKAO and the interferometric surveys are alike to HIRAX and PUMA. We implement foreground-avoidance filters and utilize models for the 21 cm thermal noise associated with single-dish and interferometer modes. The photometric galaxy surveys are similar to the DES and LSST. Our multi-tracer Fisher forecasts show a better precision for the combination of the photometric galaxy surveys and 21 cm interferometric surveys than with the 21 cm single-dish surveys - leading to at most an improvement of $23\%$ in the former case and $16\%$ in the latter case. Furthermore, we examine the impact of varying the foreground filter parameter, redshift range and sky area on the derived constraint. We find that the $f_{\rm NL}$ constraint is highly sensitive to both the redshift range and sky area. The foreground filter parameter shows negligible effect.
}

\begin{document}
\maketitle
\date{\today}
\flushbottom

\section{Introduction}
In the standard model of cosmology, Inflation leads to the generation of primordial density fluctuations, which serve as the seeds for the large-scale structure observed today.
For the simplest slow-roll single-field Inflation models, the primordial fluctuations are nearly Gaussian,
but in
many other Inflation models, fluctuations  exhibit non-Gaussianity, which leaves distinct imprints on the cosmic microwave background (CMB) and the large-scale structure. These imprints provide a powerful probe of the primordial Universe
\citep{Maldacena:2002vr, Byrnes:2010em,Celoria:2018euj,Meerburg:2019qqi}.
Local Primordial Non-Gaussianity (PNG) peaks in the squeezed configurations and thereby leaves an imprint, via scale-dependent bias, in the power spectrum of biased tracers of the dark matter \cite{Dalal:2007cu,Matarrese:2008nc}.
The strongest current constraints on the local PNG parameter $\fnl$ are from the CMB bispectrum measured by the Planck survey: $\fnl=-0.9\pm 5.1$ at the $68 \%$ confidence level \cite{Planck:2019kim}. This precision is much greater than recent galaxy survey measurements (e.g. \cite{Rezaie:2023lvi}).
Future  CMB surveys (e.g. \citep{SimonsObservatory:2018koc, CMB-S4:2016ple}), will improve this precision, but not enough to achieve a constraint on the uncertainty of the non-Gaussianity parameter, denoted $\sigma(\fnl)<1$, which would allow us to rule out a large body of Inflation models \cite{dePutter:2016trg}. 

Upcoming galaxy surveys (e.g. \cite{DESI:2022lza,Euclid:2024yrr}) and neutral hydrogen (HI) intensity mapping surveys (e.g. \cite{SKA:2018ckk,Crichton:2021hlc,PUMA:2019jwd}) will encompass many more modes than CMB surveys, opening up the prospect of higher precision measurements of $\fnl$. 
However, using only the power spectrum of a single tracer is unlikely to achieve $\sigma(\fnl)<1$ because of the growth of cosmic variance on ultra-large scales \citep{Alonso:2015uua,Raccanelli:2015vla}. 
A work-around to this obstacle is the multi-tracer method, which suppresses cosmic variance \citep{Seljak:2008xr, McDonald:2008sh,Ferramacho:2014pua,Yamauchi:2014ioa,Alonso:2015sfa,Fonseca:2015laa,Fonseca:2016xvi,Fonseca:2018hsu,Witzemann:2018cdx,SKA:2018ckk,Ballardini:2019wxj,Ginzburg:2019xsj,Gomes:2019ejy,Viljoen:2021ypp,Viljoen:2021ocx,Casas:2022vik,Jolicoeur:2023tcu,Barreira:2023rxn,Sullivan:2023qjr,Squarotti:2023nzy,Kopana:2023uew}.

In our previous work \citep{Kopana:2023uew} we used multi-tracer Fourier power spectra to combine the HI intensity maps from interferometric surveys similar to those expected from HIRAX \cite{Crichton:2021hlc} and PUMA \citep{PUMA:2019jwd}, with spectroscopic galaxy surveys similar to those expected from Euclid \cite{Euclid:2024yrr} and MegaMapper \citep{Schlegel:2022vrv}. This paper is a follow-up where we use the same HI intensity mapping surveys but also include single-dish surveys similar to those expected from  MeerKAT \cite{Santos:2017qgq} and SKA \cite{SKA:2018ckk} -- and we combine these HI intensity maps with photometric galaxy samples similar to those expected from DES \cite{DES:2021wwk} and LSST \cite{LSST:2008ijt}. 
Although photometric surveys lose accuracy in redshift space, this is less important for PNG constraints than the higher number densities  in the photometric case \cite{Alonso:2015uua,Alonso:2015sfa,Fonseca:2015laa,SKA:2018ckk}.

\section{Galaxy and intensity power spectra}\label{Multi-tracer power spectra}

For a dark matter tracer $A$, the clustering bias in the presence of local PNG  is
\begin{equation}
{b^{\rm nG}_A}(z,k) = b_{A}(z) + b_{A\phi}(z)\frac{f_{\rm NL}}{\mathcal{M}(z,k)}\,, \qquad \mathcal{M}(z,k) = \frac{2}{3\Omega_{m0}H_{0}^{2}}\frac{D(z)}{g_{\rm in}}T(k)\,k^{2}\,, \label{e2.3}
\end{equation}
where $b_A$ is the Gaussian clustering bias.
The primordial non-Gaussian bias, following \citep{Adame:2023nsx}, is assumed to be
\begin{equation}
b_{A\phi}(z) = 2 \delta_{\rm c}\big[b_{A}(z)-p_{\rm{A}}\big]\,,\quad p_A=0.955 \,,
\label{e2.4}
\end{equation}
where the critical matter overdensity is $\delta_{\rm c} =1.686$. 
(For further discussion of $b_{A\phi}$, see e.g. \citep{Barreira:2020kvh, Barreira:2021dpt, Barreira:2022sey, Barreira:2023rxn, Fondi:2023egm}.) 
In \eqref{e2.3}, $T(k)$ is the matter transfer function (normalized to 1 on very large scales), $D$ is the growth factor (normalized to 1 today) and $g_{\rm in}=(1+z_{\rm in})D_{\rm in}$ is the initial growth suppression factor, with $z_{\rm in}$ deep in the matter era. In $\Lambda$CDM with Planck best-fit parameters \cite{Planck:2018vyg}, $g_{\rm in} = 1.274$.

Then the galaxy number density  contrast, $\delta_g=\delta n_g/\bar n_g$, or the HI brightness temperature contrast, $\delta_{\rm HI}=\delta T_{\rm HI}/\bar T_{\rm HI}$, can be expressed in the general form
\begin{equation}
\delta_A(z, \bm{k}) = E_A (z,\bm{k})\Big[ {b^{\rm nG}_A}(z, k)+f(z)\mu^{2}\Big]\delta(z, \bm{k})\;,\label{e2.5}  
\end{equation}
where $f(z) = -{\ud\ln D(z)}/{\ud\ln(1+z)}$ is the growth rate, $\mu = \bm{\hat{k}} \cdot \bm{{n}}$, $\delta$ is the matter density contrast and $E_A$ is a tracer-dependent factor. For galaxies, $E_g$ accounts for photometric redshift errors, 
\begin{align}
E_{g}  = \exp\bigg\{-\frac{k^{2}\mu^{2}\big[\sigma_{z0}(1+z)\big]^{2}}{H^{2}}\bigg\} \;,\label{e2.6}
\end{align}
where $\sigma_{z0}$ is the root-mean-square (RMS) redshift error of the galaxy survey at $z=0$  \citep{Seo:2003pu,Zhan:2005ki,Green:2023uyz}.

\begin{figure}
\centering 
\includegraphics[width=7.6cm]{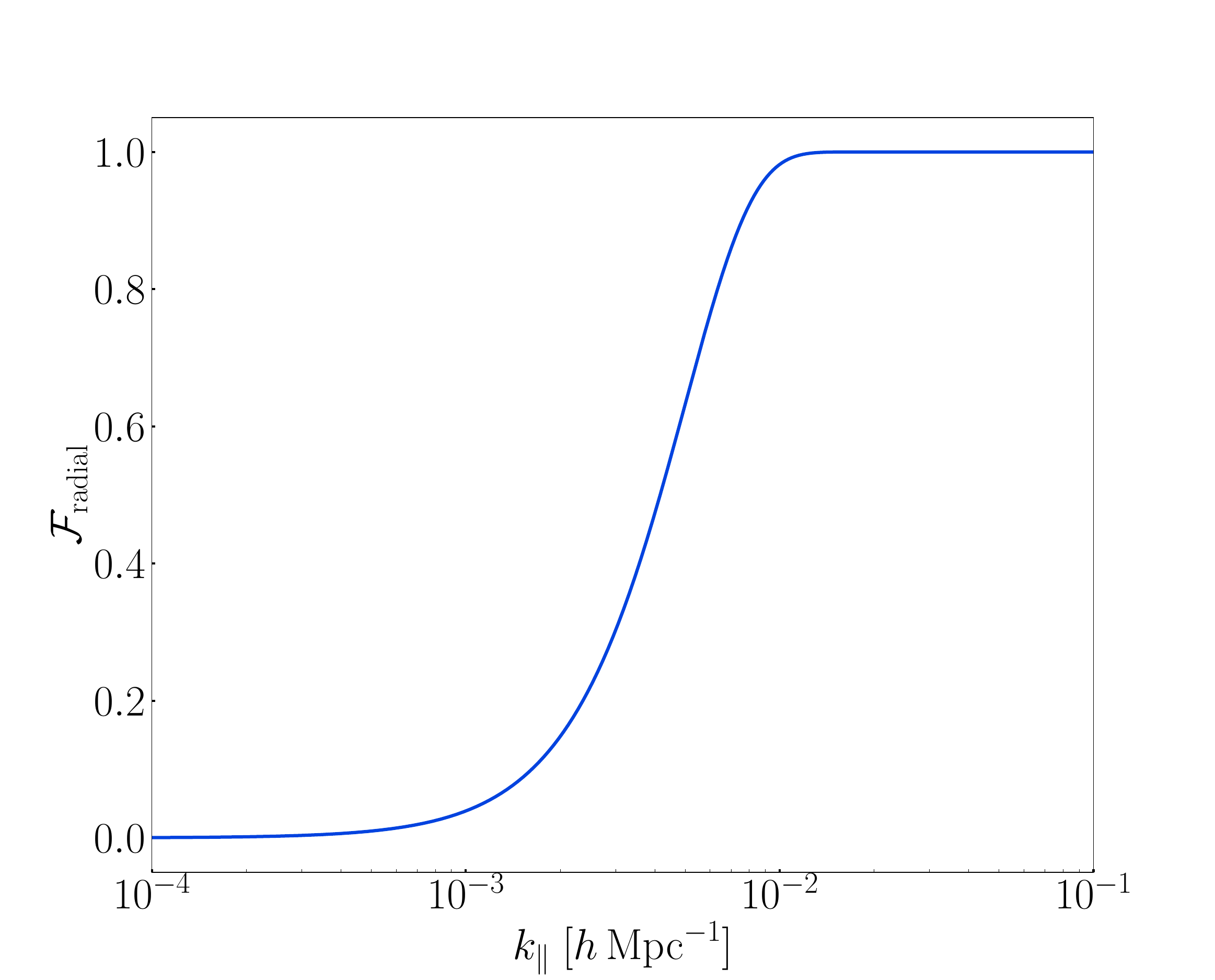} 
\includegraphics[width=7.6cm]{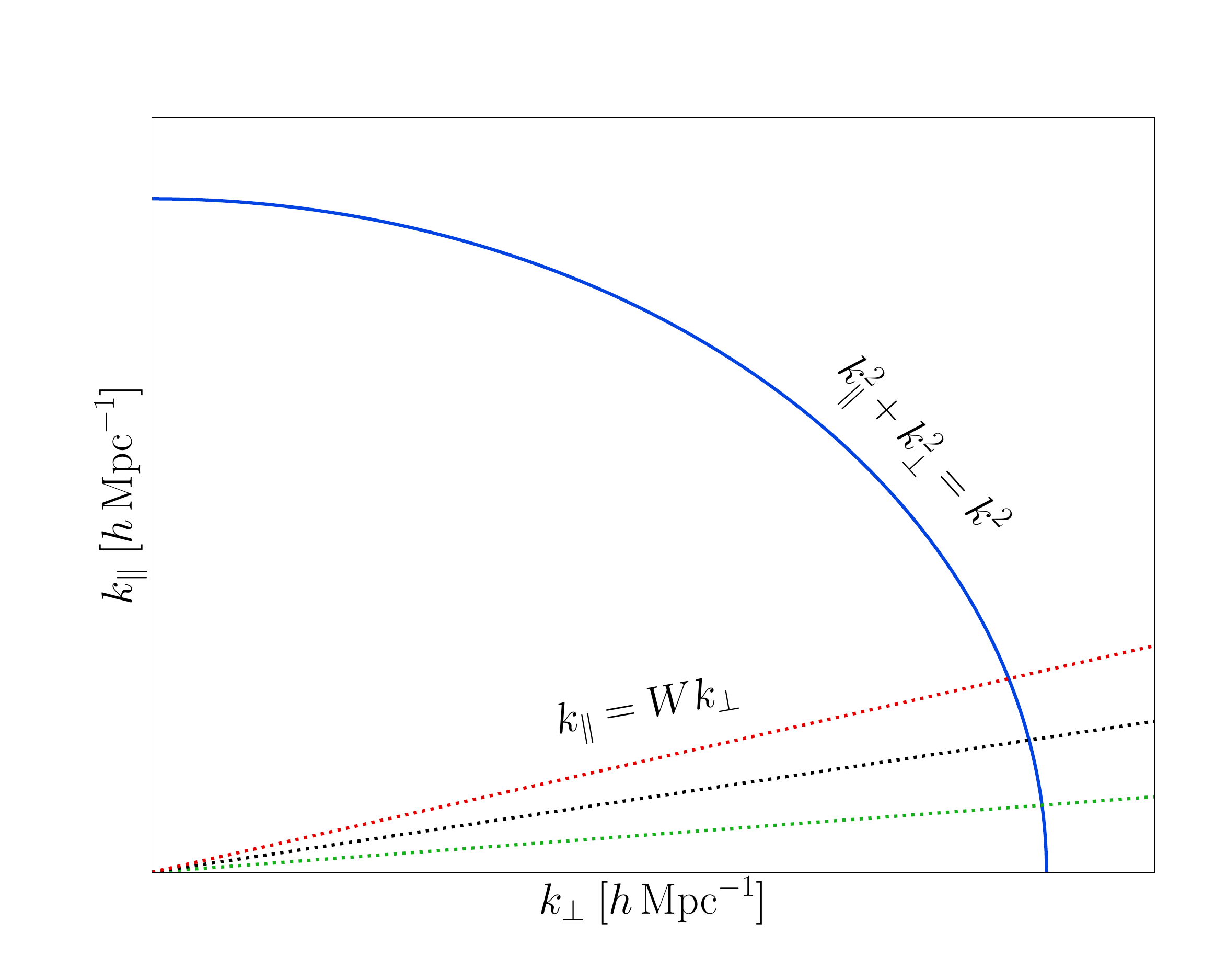} 
\caption{For HI intensity mapping: radial foreground damping function  (\emph{left}); interferometer foreground wedge region, whose size increases with redshift (\emph{right}). 
}\label{fig6}
\end{figure}

For HI intensity, $E_{\rm HI}$ accounts for foreground avoidance and telescope beam effects:

\begin{align}
E_{\rm HI}=  {\cal F}{\cal B}\;,  \label{e3.19} 
\end{align}
where the foreground and beam factors depend
on whether the HI intensity map is made in interferometer or single-dish mode. In the case of the foreground factor:
\begin{align}
{\cal F}= {\cal F}_{\rm radial}\,{\cal F}_{\rm wedge}=
\bigg[1 - \exp\bigg\{-\mu^2\bigg(\frac{k} {k_{\rm fg}}\bigg)^{\!\!{2}\,}\bigg\}\bigg]{\cal F}_{\rm wedge}\,.\label{cfcw}
\end{align}
The radial damping term affects both single-dish and interferometer surveys. It models the loss of large-scale radial wave-modes due to foreground contamination \citep{Cunnington:2020wdu,Cunnington:2021czb,Spinelli:2021emp,Cunnington:2023jpq}. We use $k_{\rm fg}=0.005$ and 0.01\,$h/$Mpc. 
The term ${\cal F}_{\rm wedge}$ is 1 for single-dish surveys. For interferometer surveys,
there is additional foreground contamination in a wedge-shaped region of $\bm k$-space defined by \cite{Pober:2013jna,Pober:2014lva,Obuljen:2017jiy,Alonso:2017dgh,Ghosh:2017woo,PUMA:2019jwd,Karagiannis:2019jjx}:
\begin{align}
\left|k_{\parallel}\right| &< W(z) \, k_{\perp}\;, \label{e3.20}
\\
W(z) &= r(z) \mathcal{H}(z) \sin\big[0.61 N_{\mathrm{w}} \,\theta_{\mathrm{fov}}(z)\big] \;. \label{e3.21}
\end{align}
Here $k_{\parallel}=\mu k$, $k_{\perp}=(1-\mu^2)^{1/2}k$, $r$ is the comoving radial distance, $\cH=H/(1+z)$ is the conformal Hubble rate and
 {$\theta_{\mathrm{fov}} = 1.22\,{\lambda_{21}(1+z)}/{D_{\rm d}}$} is the field of view, with $D_{\rm d}$ the dish diameter.
The extent of foreground contamination is determined by the number of primary beams affected, $N_{\rm{w}}$. 
We choose $N_{\mathrm{w}}=1$.
For interferometer surveys, ${\cal F}_{\rm wedge}$  in \eqref{cfcw} is the Heaviside step function,
\begin{align}
{\cal F}_{\rm wedge} =\Theta\big(\left|k_{\parallel}\right| - W(z)\, k_{\perp} \big) .   
\end{align}
\autoref{fig6} illustrates ${\cal F}_{\rm radial}$ and the wedge region \eqref{e3.21}.
The beam factor ${\cal B}$ in \eqref{e3.19} depends on whether the HI intensity map is made in interferometer or single-dish mode:
\begin{align}
\text{interferometer mode:}\quad & {\cal B} = \theta_{\rm fov} \\  
\text{single-dish mode:}\quad &  {\cal B} = \exp\bigg(-\frac{k_{\perp}^{2}r^{2}\theta_{\rm fov}^2}{16\ln 2}\bigg)  \;,  
\end{align}

The power spectra are given by 
\begin{equation}
\big \langle \delta_A(z,\bm{k})\delta_{B}(z,\bm{k}') \big \rangle = (2\pi)^{3}P_{AB}(z,\bm{k})\delta^{\mathrm{D}}(\bm{k}+\bm{k}')\;, \label{e2.7}
\end{equation}
so that
\begin{equation}
P_{AB}=E_{A}E_{B}\big( {b^{\rm nG}_A}+f\mu^2 \big)\big( {b^{\rm nG}_B}+f\mu^2 \big)P
\;,\label{e2.8}  
\end{equation}
where $P$ is the matter power spectrum (computed with CLASS \citep{Blas:2011rf}). When $B=A$ we use the shorthand $P_A \equiv P_{AA}$.


\section{Galaxy and intensity mapping surveys}\label{Surveys}

\begin{table}  
\caption{Specifications of mock surveys. Here D, L, M, S, H, P have properties similar to DES, LSST, MeerKAT (UHF Band), SKA (Band 1), HIRAX, PUMA respectively.
 }

\centering 
\label{tab1} 
\vspace*{0.4cm}

\begin{tabular}{ c| c c c} 
\hline\hline   \\ [-0.8ex]
  HI single-dish $\otimes $ Photometric & $z$ & $\Omega_{\rm sky}~[{\rm deg}^2]$ &  $t_{\rm tot}~[\rm{hr}]$
\\ [0.8ex]
\hline  
 {M} $\otimes $ D   & $0.4-1.45$    & $5,000$   & $4,000$  \\[1.0ex] 
 {S} $\otimes$ D      & $0.35-2.0$     & $5,000$   & $2,500$  \\[1.0ex]
  {S} $\otimes$ L     & $0.35-2.9$     & $10,000$  & $5,000$ \\[2.0ex]\hline\hline
 
 HI interferometer $\otimes$  Photometric  & &\\[1.0ex]\hline
H $\otimes$ D           & $0.8-2.0$     & $5,000$   & $5,833$ \\[1.0ex]
H $\otimes$ L           & $0.8-2.5$     & $10,000$  & $12,000$\\[1.0ex]
P $\otimes$ L          & $0.3-2.9$     & $10,000$  & $19,000$ \\[2.0ex]\hline\hline
\end{tabular} 
\end{table}

In this paper, our aim is not to produce forecasts for specific surveys, but rather to estimate the improvements in precision that are possible with a multi-tracer analysis. To this end, we consider mock surveys which combine very different pairs of tracers: a photometric galaxy sample and an HI intensity map.  The basic properties assumed for the mock surveys are similar to those of surveys considered in the literature: 
\begin{itemize}
    \item 
    Photometric galaxy samples D and L, similar to those from DES (blue and red) \cite{DES:2021wwk} and LSST (blue) \cite{LSST:2008ijt} respectively.
    \item 
    HI intensity mapping samples:
    \begin{itemize}
    \item single-dish mode samples M and S, similar to those from  MeerKAT (UHF Band) \cite{Santos:2017qgq} and SKA-MID (Band~1) \cite{SKA:2018ckk} respectively;
    \item  interferometer mode samples H and P, similar to those from HIRAX \cite{Crichton:2021hlc} and PUMA \cite{PUMA:2019jwd} respectively.
    \end{itemize}
\end{itemize}

\autoref{tab1} summaries the basic survey specifications that we assume. Note that for each multi-tracer pair and for the single tracers in the pair, we only consider the overlapping sky area and redshift range -- since our focus is on the improvement delivered by the multi-tracer analysis. In the case of future surveys we assume a nominal overlap sky area of 10,000\,deg$^2$.

The Gaussian clustering bias for galaxies and intensity maps is assumed known up to an overall amplitude parameter, following \cite{Agarwal:2020lov}:
\begin{equation}
    b_{A}(z) = b_{A0}\, {\beta_A}(z) \label{e2.2}\,.
\end{equation}
For the photometric D and L samples we assume
\begin{align}
b_{\rm D}(z) &= b_{\rm D 0}\big(1 +  0.84 z \big) & \text{fiducial}~ b_{\rm D 0}=1.0~ \label{e3.3}   ~~\citep{Fonseca:2016xvi},\\
b_{\rm L}(z)&= b_{\rm L 0}/D(z) & \text{fiducial}~ b_{\rm L 0}=0.95 \label{e3.2} ~~\citep{Alonso:2015sfa}.
\end{align}
For the photometric surveys we assume
\begin{equation}
    \sigma_{z0}  = 0.05\;.\label{e3.1}
\end{equation}
All HI intensity surveys have the same Gaussian clustering bias. Following \citep{Villaescusa-Navarro:2018vsg} we assume
\begin{align}
b_{\rm HI}(z)&= b_{\rm HI 0}\big(1 + 0.823 z - 0.0546 z^{2}\big) & \text{fiducial}~ b_{\rm HI 0}=0.842 ~.\label{e3.6} 
\end{align}

For the L sample, we use the background comoving number density for LSST given in Table 8 of \citep{Green:2023uyz}. For the D sample, we use the fitting formula for the DES number per redshift per solid angle \citep{DES:2015bqp,Fonseca:2016xvi}:
 \begin{align}
     \bar{N}_{\rm D}(z) = 8.05\times10^4\Bigg{(}\frac{z}{0.57}\Bigg{)}^{1.04}  \exp\bigg[-\Big( \frac{z}{0.57}\Big)^{\!1.34}\bigg]~~\mathrm{deg}^{-2}\,. \label{e3.5}
 \end{align} 
Then the comoving number density is given by
\begin{equation}
    \bar{n}_{\rm D} = \frac{(1 + z)\mathcal{H}}{ {r}^{2}} \,\bar{N}_{\rm D}\;.\label{e3.4}
\end{equation}

The background brightness temperature is the same for all HI intensity samples and we  model it as \citep{Santos:2017qgq}:
\begin{equation}
\bar{T}_{\rm HI}(z) = 0.0559 +0.2324\,z -0.0241\, z^{2} ~~ \mathrm{mK}\,. \label{e3.7}
\end{equation} 

\begin{figure}
\centering
\includegraphics[width=7.5cm]{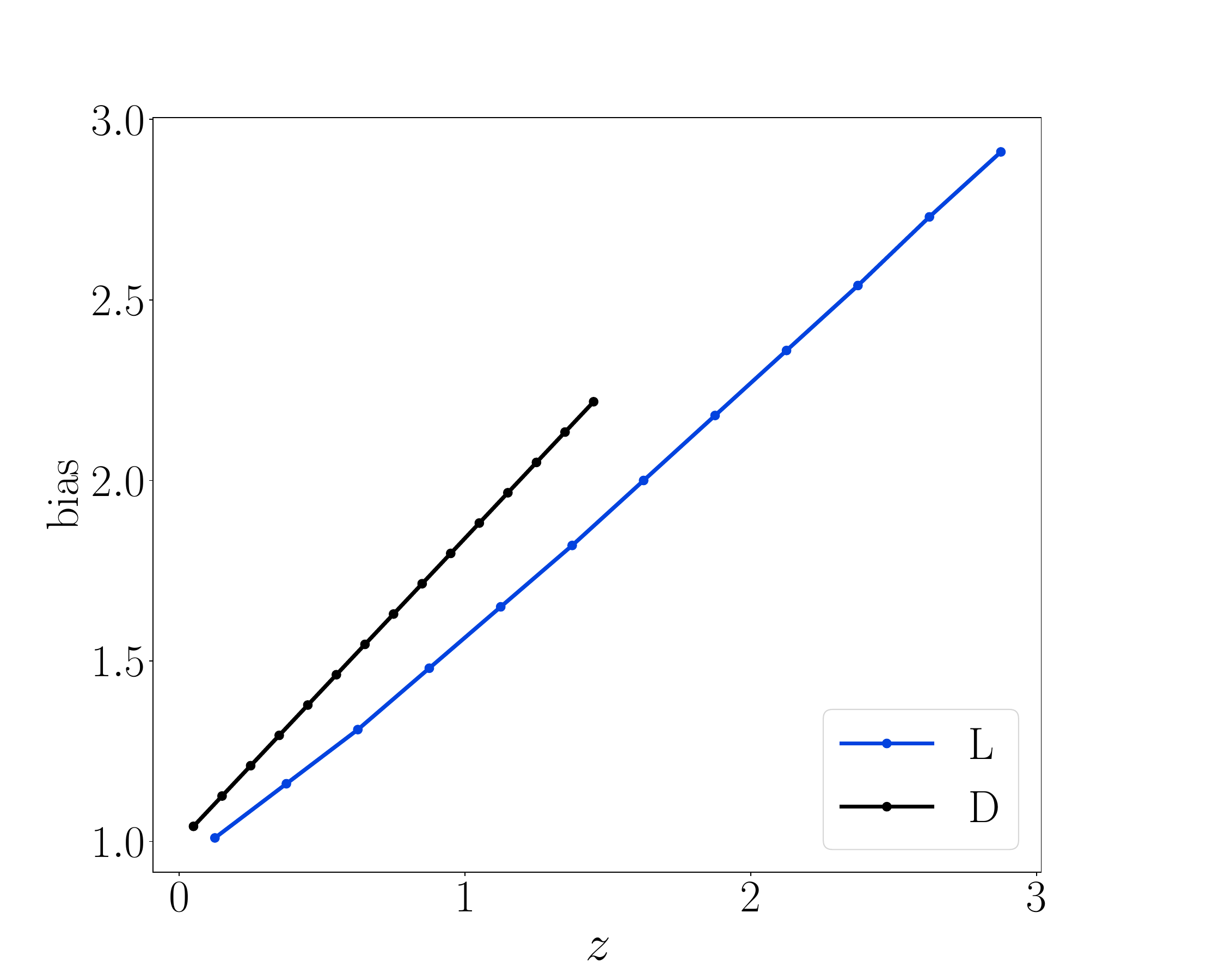} 
\includegraphics[width=7.5cm]{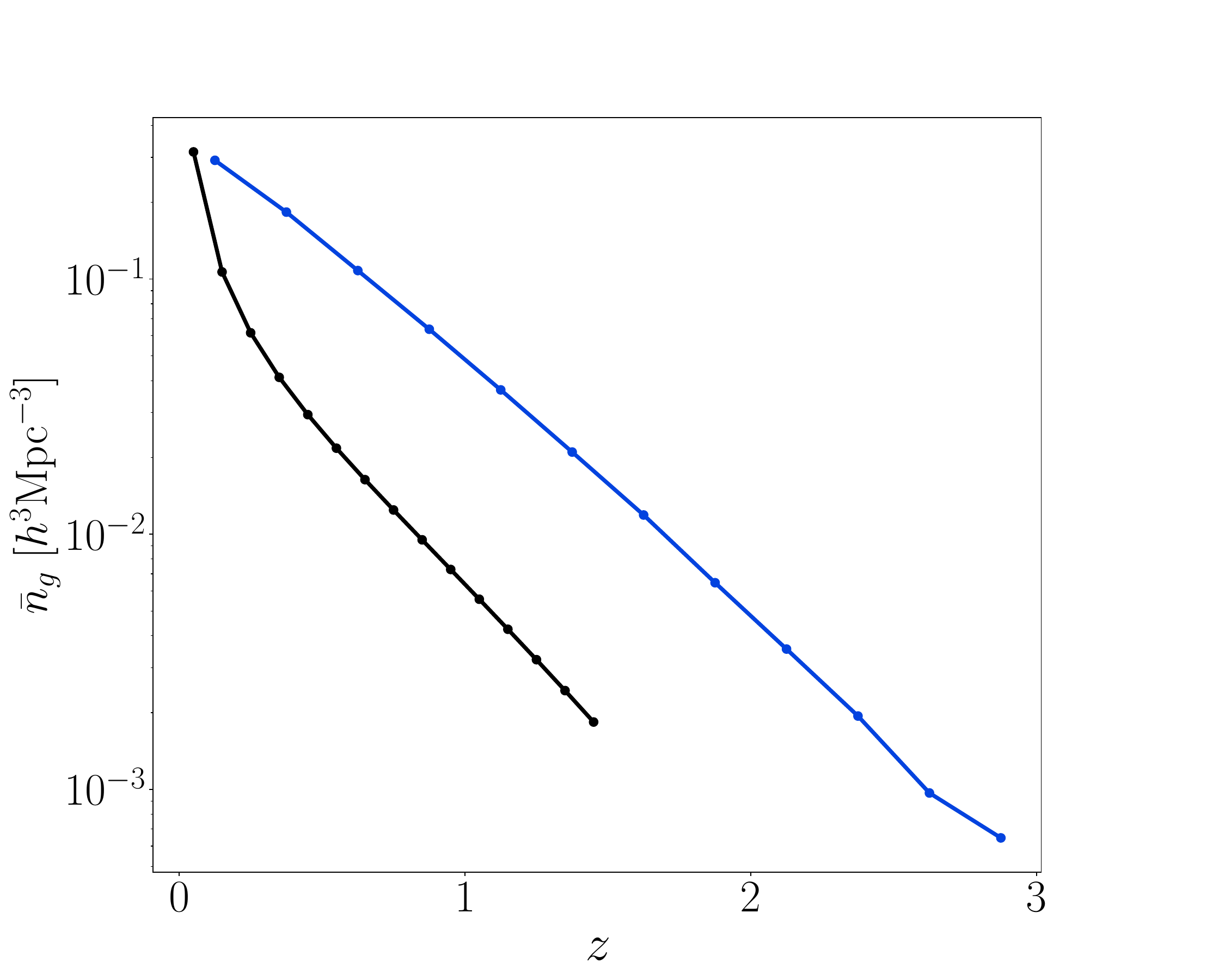}  
\caption{Galaxy surveys: Gaussian clustering biases ({\em left}); 
comoving number densities ({\em right}).} 
\label{fig1}
\end{figure}

\begin{figure}
\centering 
\includegraphics[width=7.5cm]{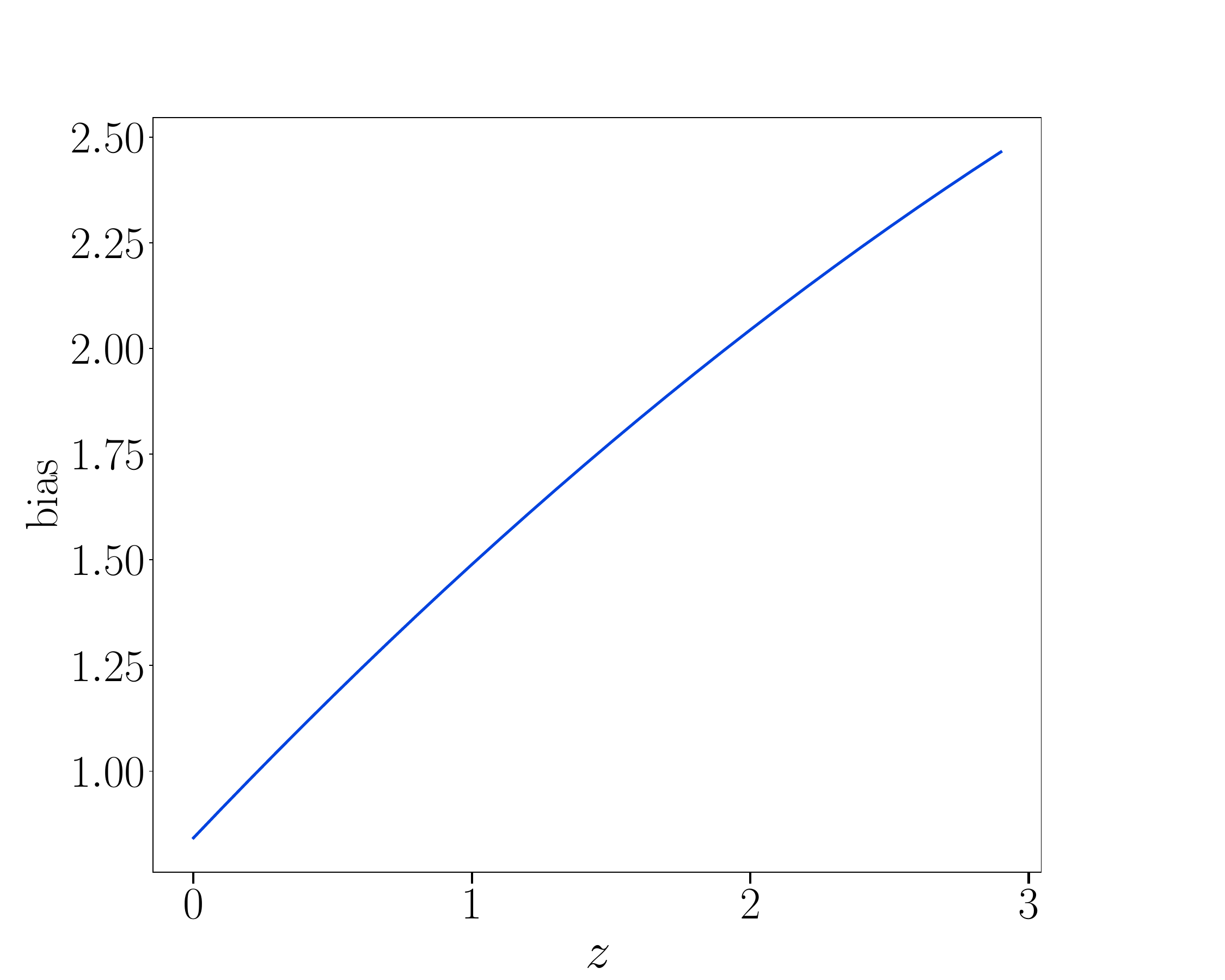} 
\includegraphics[width=7.5cm]{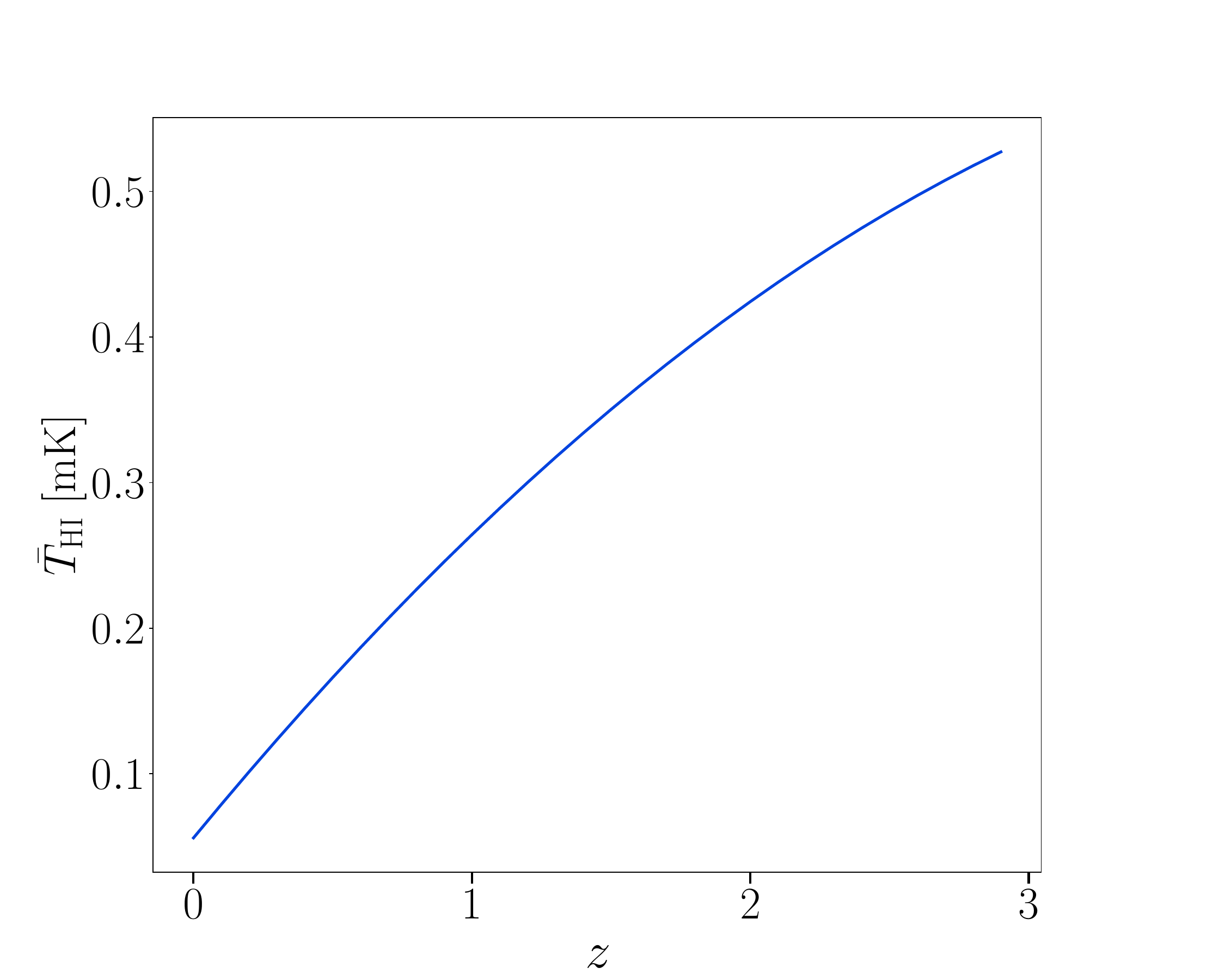}
\caption{HI intensity maps: Gaussian clustering bias ({\em left}); background brightness temperature ({\em right}).
} 
\label{fig2}
\end{figure}

The Gaussian clustering biases  and number densities for L (LSST-like) and D (DES-like) are displayed in  \autoref{fig1}.  \autoref{fig2} displays the  HI intensity Gaussian bias and the background brightness temperature.
 

\section{Noise power spectra}\label{Noise}
The measured tracer power spectrum is made up of the cosmological signal and the noise, so that the total observed auto-power is
\begin{equation}
\tilde{P}_{A} = P_{A} +  P_{A\,\rm noise}\;. \label{e3.8}
\end{equation}
For galaxy surveys, $A=g$, the noise power spectrum is dominated by shot noise:
\begin{equation}
P_{g\, \rm noise} =  \frac{1}{\bar n_{ g}}\;. \label{e3.10}
\end{equation}
For HI intensity mapping surveys, $A=\rm{HI}$,  on the linear scales that we consider, the noise power is dominated by the thermal noise and shot noise can be neglected \cite{Castorina:2016bfm, Villaescusa-Navarro:2018vsg}. 
The shot noise cross-power between HI intensity and galaxy surveys can also be neglected \cite{Viljoen:2020efi,Casas:2022vik}. Since HI thermal noise does not correlate with galaxy shot noise, we neglect the noise cross-power:
\begin{equation}
 {\tilde P_{g \rm HI}=P_{g \rm HI}}\,. \label{e3.9}
\end{equation}

For HI intensity mapping surveys, the thermal noise power spectrum is \cite{Bull:2014rha,Alonso:2017dgh,PUMA:2019jwd,Jolicoeur:2020eup}:
\begin{equation}
P_{\rm HI\,  noise} = \frac{\lambda_{21}\Omega_{\mathrm{sky}}}{2 t_{\mathrm{tot}}}\,\frac{(1+z) r^{2} }{\cH }\,\left(\frac{T_{\mathrm{sys}}}{\bar{T}_{\rm HI}}\right)^2\,
\alpha 
\;, 
\label{e3.11}
\end{equation}
where we used natural units (speed of light = 1) and
\begin{align}
 \text{single-dish mode:}\quad &  \alpha = \frac{1}{N_{\rm d}}\;, \label{3.12}\\
 \text{interferometer mode:}\quad &        \alpha = \left[\frac{4\lambda_{21}(1+z)}{0.7\pi D_{\mathrm{d}}^{2}}\right]^2\frac{1}{n^{\mathrm{phys}}_{\mathrm{b}}}\;. \label{3.13}
    \end{align}
Here $N_{\rm d}$ is the number of dishes and the system temperature is  \citep{PUMA:2019jwd}:
\begin{equation}
T_{\mathrm{sys}} = T_{\rm d}+T_{\rm sky} =T_{\rm d} + 2.7 + 25\bigg[\frac{400\,\mathrm{MHz}}{\nu_{21}} (1+z)\bigg]^{2.75} ~ \mathrm{K},\label{e3.18}
\end{equation} 
where $T_{\rm d}$ is the dish receiver temperature. All specifications are given in \autoref{tab2}.

\begin{table} 
\centering 
\caption{Specifications of HI intensity mapping surveys:   {M (MeerKAT-like), S (SKA-like), H (HIRAX-like), P (PUMA-like). 
$N_{\rm d}$ is the number of dishes, $D_{\rm d}$ is the dish diameter, $T_{\rm d}$ is the dish receiver temperature, $D_{\rm max}$ is the maximum baseline (for interferometer surveys), $t_{\rm tot}$ is the observing time. The interferometer parameters $N_{\rm s}$ and $a,\cdots, e$ appear in \eqref{e3.15}.}
}~\\
 \label{tab2} 
\vspace*{0.2cm}
\begin{tabular}{|l|c|c|c|c|c|} 
\hline 
Survey~~&~~$N_{\mathrm{d}}$~~& ~~$N_{\mathrm{s}}$~~&~~$T_{\mathrm{d}}$~~& ~~$D_{\mathrm{d}}$~~& ~~$D_{\mathrm{max}}$~~ \\ 
 & & &~~[K]~~&~~[m]~~&~~[m]~~ \\ \hline \hline
  {M} & 64 & - & 25 & 13.5 & - ~~ \\
  {S} & 197 & - & 25 & 15 & - ~~ \\
H(256,\,1024) & 256,\,1024 & 16,\, 32 & 50 & 6 & 141,\, 282~~ \\
P(5k,\,32k) & 5k,\,32k &  100,\,253 & 93 & 6 & 648,\,1640\\ \hline \\ \hline 
Fitting &~~$a$~~ &~~$b$~~ &~~$c$~~&~~$d$~~&~~$e$ \\
parameters & & & & & \\
\hline \hline
H(256,\,1024) & 0.4847 & -0.3300 & 1.3157 & 1.5974 & 6.8390\\ 
P(5k,\,32k) & 0.5698 & -0.5274 & 0.8358 & 1.6635 & 7.3177  \\ \hline 
\end{tabular}
\end{table}

\begin{figure}
\centering
\includegraphics[width=7.5cm]{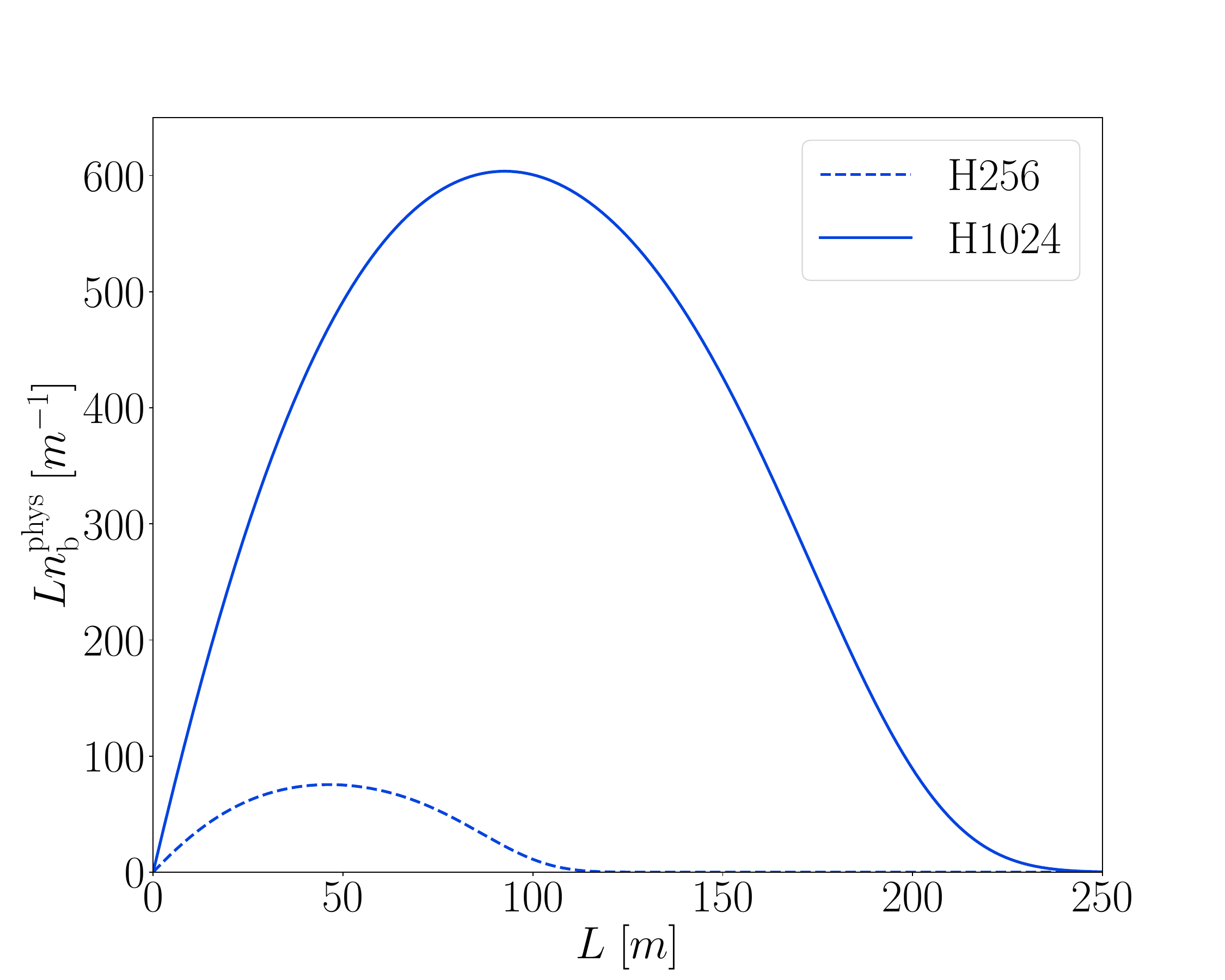} 
\includegraphics[width=7.5cm]{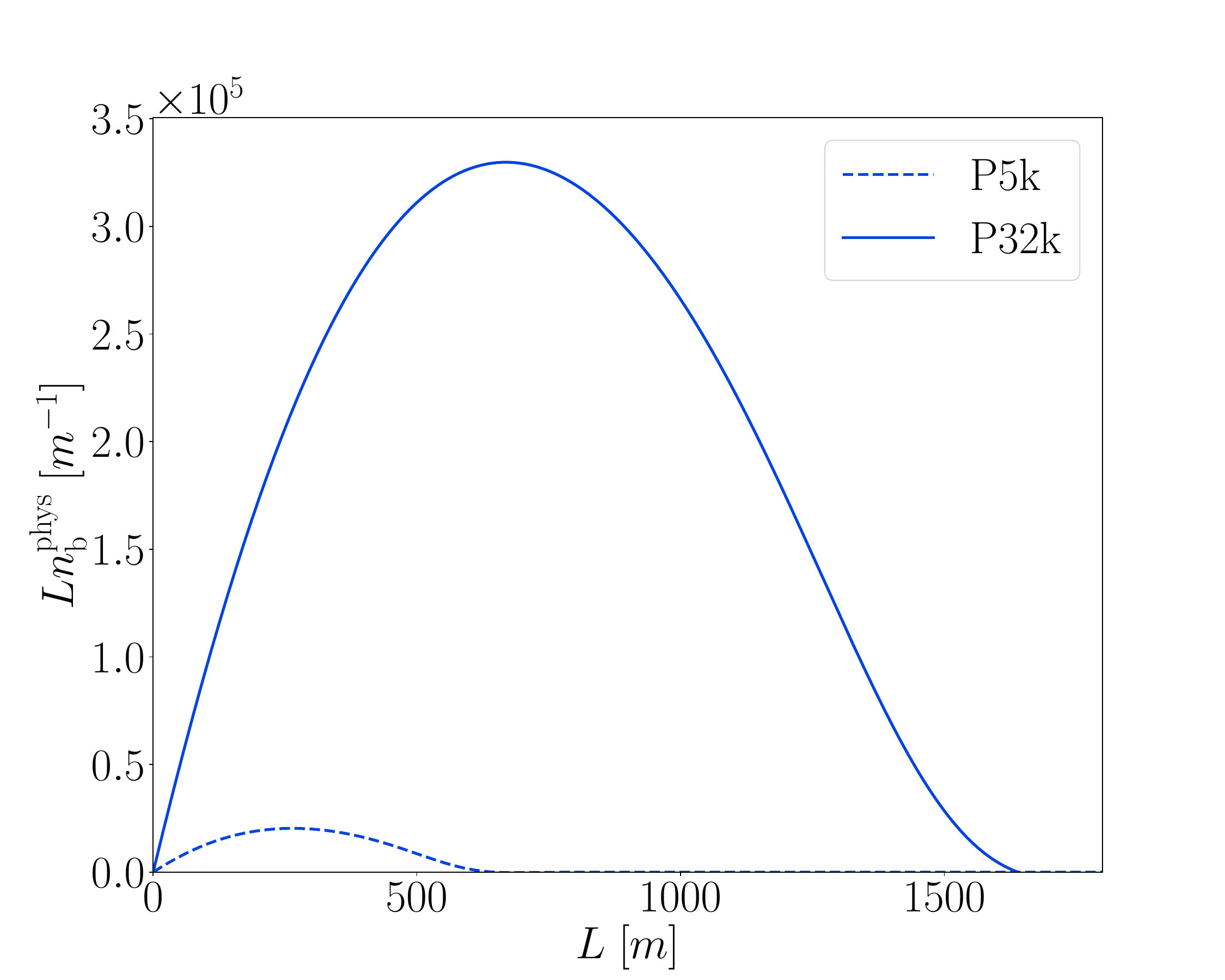} 
\caption{ Physical baseline density for HI IM surveys:  {HIRAX-like  (\emph{left}) and PUMA-like  (\emph{right})}.
} \label{fig5}
\end{figure}

The physical baseline density of the  HIRAX- and PUMA-like surveys is modelled following
\cite{PUMA:2019jwd}:
\begin{equation}
n^{\mathrm{phys}}_{\mathrm{b}} = \left(\frac{N_{\mathrm{s}}}{D_{\mathrm{d}}}\right)^{2}\frac{a+b\,(L/L_{\mathrm{s}})}{1+c\,(L/L_{\mathrm{s}})^{d}}\exp{\big[-(L/L_{\mathrm{s}})^{e}\big]}\,,\quad 
L = \frac{k_{\perp}r}{2\pi}\,\lambda_{21}(1+z)\;.
\label{e3.15}
\end{equation}

H (HIRAX-like) is a square close-packed array, whereas P (PUMA-like) is a hexagonal close-packed array with a $50\%$ fill factor -- meaning that dish sites occupy half of the total area  \citep{PUMA:2019jwd,Karagiannis:2019jjx}. Then it follows that $N_{\rm s}^2 =N_{\rm d}$ for H and $N_{\rm s}^2 =2N_{\rm d}$ for P \citep{PUMA:2019jwd, Castorina:2020zhz}, where $L_{\mathrm{s}}=N_{\mathrm{s}}D_{\mathrm{d}}$. The values of the fitting parameters $a,b,c,d,e$ in \eqref{e3.15} are given in \autoref{tab2}, from \cite{PUMA:2019jwd} (their Appendix D, last equation). See \autoref{fig5} for plots of the baseline density.

\newpage
\section{Fisher forecast}\label{Fisher forecast}

Three cosmological and two nuisance parameters,
\begin{equation}
\vartheta_{\alpha} = \big\{f_{\mathrm{NL}},\;A_{s},\;n_{s},\;b_{g0},\;b_{{\rm HI} 0} \big\}\,,\label{e4.1}
\end{equation}
are considered, with fiducial values for the LCDM cosmological parameters given by the  {\em Planck\;2018} best-fit values \cite{Planck:2018vyg}, including $\bar A_{s}=2.105\times 10^{-9}$ and $\bar n_{s}=0.9665$. We assume that the PNG fiducial is $\bar{f}_{\rm NL}=0$ and the fiducial values of $b_{Ao}$ are given in \eqref{e3.3}, \eqref{e3.2} and \eqref{e3.6}. 

The Fisher information matrix in each redshift bin is
\begin{equation}
F_{\alpha \beta}^{\bm{P}} = \sum_{\mu=-1}^{+1}\,\sum_{k=k_{\mathrm{min}}}^{k_{\mathrm{max}}}\,\partial_{\alpha}\,\bm{P} \cdot \mathrm{Cov}(\bm{P},\bm P)^{-1} \cdot \partial_{\beta}\,\bm{P}^{\mathrm{T}}\;. \label{e4.2}
\end{equation}
Here $\partial_{\alpha}  = \partial /\partial \vartheta_{\alpha}$, the data vector of power spectra is 
\begin{equation}
\bm{P} = \big( P_{g}\,,\,  P_{g{\rm HI}}\,,\,  P_{\rm HI} \big)\;,\label{e4.3}
\end{equation}
and the covariance is \cite{White:2008jy,Zhao:2020tis,Barreira:2020ekm,Karagiannis:2023lsj}: 
\begin{equation} 
\mathrm{Cov}(\bm{P},\bm P) =  \frac{k_{\rm f}^3}{2\pi k^{2} \Delta k}\,\frac{2}{\Delta \mu}\,
\begin{pmatrix}
\tilde P_{g}^2 & & \tilde P_{g}\tilde P_{g{\rm HI}} & & \tilde P_{g{\rm HI}}^2 \\ 
& & & & \\
\tilde P_{g}\tilde P_{g{\rm HI}} & & \frac{1}{2}\big(\tilde P_{g}\tilde P_{\rm HI}+ \tilde P_{g{\rm HI}}^2 \big) & & \tilde P_{\rm HI}\tilde P_{g{\rm HI}} \\
& & & & \\
\tilde P_{g{\rm HI}}^2 & & \tilde P_{\rm HI}\tilde P_{g{\rm HI}} & & \tilde P_{\rm HI}^2 
\end{pmatrix}\;. \label{e4.4}
\end{equation} 
We choose
\begin{equation}
 \Delta z=0.1\,,\quad \Delta \mu = 0.04\,, \quad \Delta k = k_{\mathrm{f}}={2\pi}{V^{-1/3}}\,,\quad 
 k_{\mathrm{max}} = 0.08\big(1+z\big)^{2/(2+n_{s})}\,h/{\rm Mpc}\,,
 \label{e4.5}
\end{equation}
where $V$ is the volume of the bin and
\begin{align}
A=g:~~k_{\rm min}=k_{\rm f}\,,\qquad
A={\rm HI}:~~k_{\rm min}={\rm max}\, \big(k_{\rm f},\, k_{\rm fg} \big)\,.
\end{align}

\section{Results}

\begin{table}[h]  
\caption{Marginalised $68\%$ CL errors on $\fnl$ from galaxy surveys D (DES-like), L (LSST-like) and HI IM single dish-mode surveys A = M (MeerKAT UHF-like), S (SKA Band 1-like).  
} 
\centering 
\label{tab3} 
\vspace*{0.4cm}

\small

\begin{tabular}{ccc}
 M $\otimes$ D  &  S $\otimes$ D  & S $\otimes$ L \\
\begin{tabular}{c r r} 
\hline\hline   \\ [-0.8ex]
  Survey $ A, A\otimes B$  & & $\sfnl$
\\ [0.8ex]
\hline  
 D  & &   13.0 \\[1.0ex]
    M & &  22.5  \\[1.0ex]
    M $\otimes$ D & &  10.3  \\[2.0ex]\hline\hline

\end{tabular} &

\begin{tabular}{c r r} 
\hline\hline   \\ [-0.8ex]
   
 Survey $ A, A\otimes B$   & & $\sfnl$
\\ [0.8ex]
\hline  
    D & &  6.89 \\[1.0ex]
   S & &  10.9 \\[1.0ex]
   S $\otimes$ D & & 5.78 \\[2.0ex]\hline\hline
\end{tabular} &

\begin{tabular}{c r r} 
\hline\hline   \\ [-0.8ex]
   
  $ A, A\otimes B$   & & $\sfnl$
\\ [0.8ex]
\hline  
     L &&  2.28  \\[1.0ex]
   S &&  4.15\\[1.0ex]
   S $\otimes$ L & & 2.15  \\[2.0ex]
\hline\hline 
\end{tabular}
   
\end{tabular}  
\end{table}

\begin{table}[h]  
\caption{Marginalised $68\%$ CL errors on $\fnl$ from galaxy surveys A = D (DES-like), L (LSST-like) and HI IM inteferometer-mode surveys A = H (HIRAX-like) and P (PUMA-like). H and P have phases 1 and 2 with the initial and final number of dishes.  
} 
\centering 
\label{tab4} 
\vspace*{0.4cm}

\small

\begin{tabular}{ccc}
H $\otimes$ D & H $\otimes$ L &  P $\otimes$ L\\
\begin{tabular}{c r r} 
\hline\hline   \\ [-0.8ex]
   Survey $ A, A\otimes B$ & & $\sfnl$
\\ [0.8ex]
\hline  
  D & &  7.00   \\[1.0ex]
  H256 & & 11.40  \\[1.0ex]
    H1024 & & 9.09   \\[1.0ex]
      H256 $\otimes$ D & & 5.70  \\[1.0ex]
    H1024 $\otimes$ D & & 5.40   \\[2.0ex]\hline\hline
\end{tabular} &

\begin{tabular}{c rr r} 
\hline\hline   \\ [-0.8ex]
   
   Survey $ A, A\otimes B$ & & $\sfnl$
\\ [0.8ex]
\hline  
 L  & & 2.92   \\[1.0ex]
   H256 & & 5.15   \\[1.0ex]
    H1024 & & 4.03   \\[1.0ex]
       H256 $\otimes$ L & & 2.61   \\[1.0ex]
    H1024 $\otimes$ L & & 2.51   \\[2.0ex]\hline\hline

\end{tabular} &

\begin{tabular}{c rr r} 
\hline\hline   \\ [-0.8ex]
   
  $ A, A\otimes B$ & & $\sfnl$
\\ [0.8ex]
\hline  
     L & & 2.22   \\[1.0ex]
   P5k & & 2.48  \\[1.0ex]
   P32k & & 2.28  \\[1.0ex]
   P5k $\otimes$ L & & 1.85  \\[1.0ex]
   P32k $\otimes$ L & & 1.81  \\[2.0ex]
\hline\hline 
\end{tabular}
\end{tabular}  
\end{table}  

The main results are presented in \autoref{tab3} and \autoref{tab4}, showing the errors on $\fnl$ for HI intensity mapping surveys combined with photometric surveys, after marginalising over $A_s,n_s$ and the nuisance parameters $b_{A0}$. We consider multi-tracer combinations of the form HI intensity~$\otimes$~galaxy, using 
HI  surveys  M (like MeerKAT UHF Band) and S  (like SKA Band 1) in single-dish mode, and H (HIRAX-like) and P (PUMA-like) in interferometer mode, combined with galaxy surveys D (DES-like) and L (LSST-like). 

We adjusted the HI intensity observational time in each case to be consistent with the overlapping sky area and redshift range. For each multi-tracer scenario $A \otimes B$, we used the same sky area and redshift range for the two single-tracers and the multi-tracer. We omitted the case P $\otimes$ D, because of poor overlap in redshift.  
Note that the full sky area and redshift range of each survey are not used in every multi-tracer combination -- since we only use the {\em overlappping} sky area and redshift range. This explains for example why D (DES-like) achieves better precision than S (SKA-like) -- since D in the multi-tracer is the full DES-like survey, while S is only a  fraction of the full SKA-like survey.

The results clearly demonstrate the improvement in precision on local PNG from the multi-tracer. 
At higher redshifts, such as those covered by SKA-like (\autoref{tab3}) and HIRAX- and PUMA-like (\autoref{tab4}), the improvement in precision is more notable, especially for SKA-  and PUMA-like, due to the large sky volumes probed.

We also note that the first phases of HIRAX (256 dishes) and PUMA (5k dishes) perform nearly as well in the multi-tracer as the second phases. A high number of dishes for these interferometers is not necessary for $\fnl$ constraints.

\begin{figure}[!htbp]
\centering
\includegraphics[width=7.5cm]{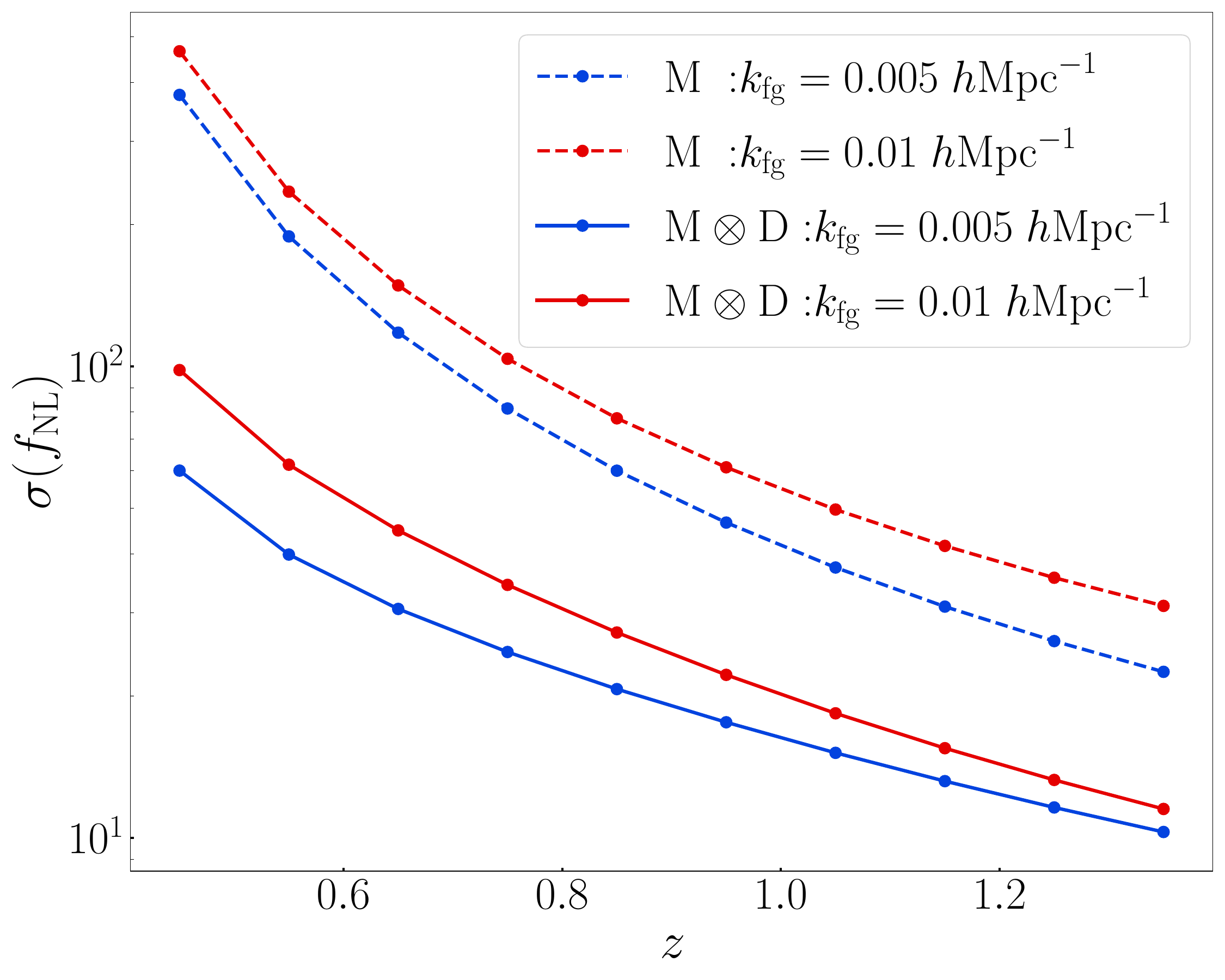}\\
\includegraphics[width=7.5cm]{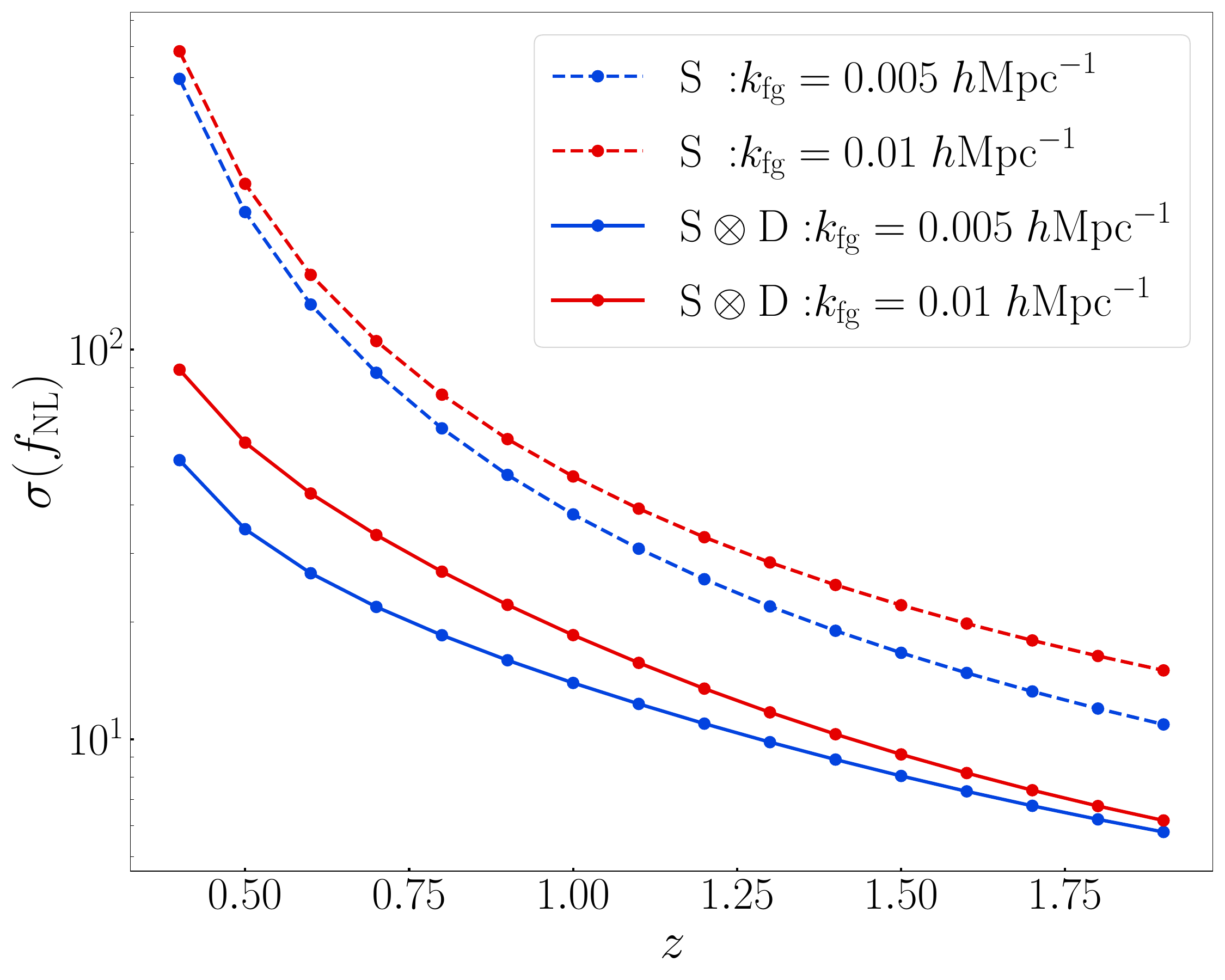}
\includegraphics[width=7.5cm]{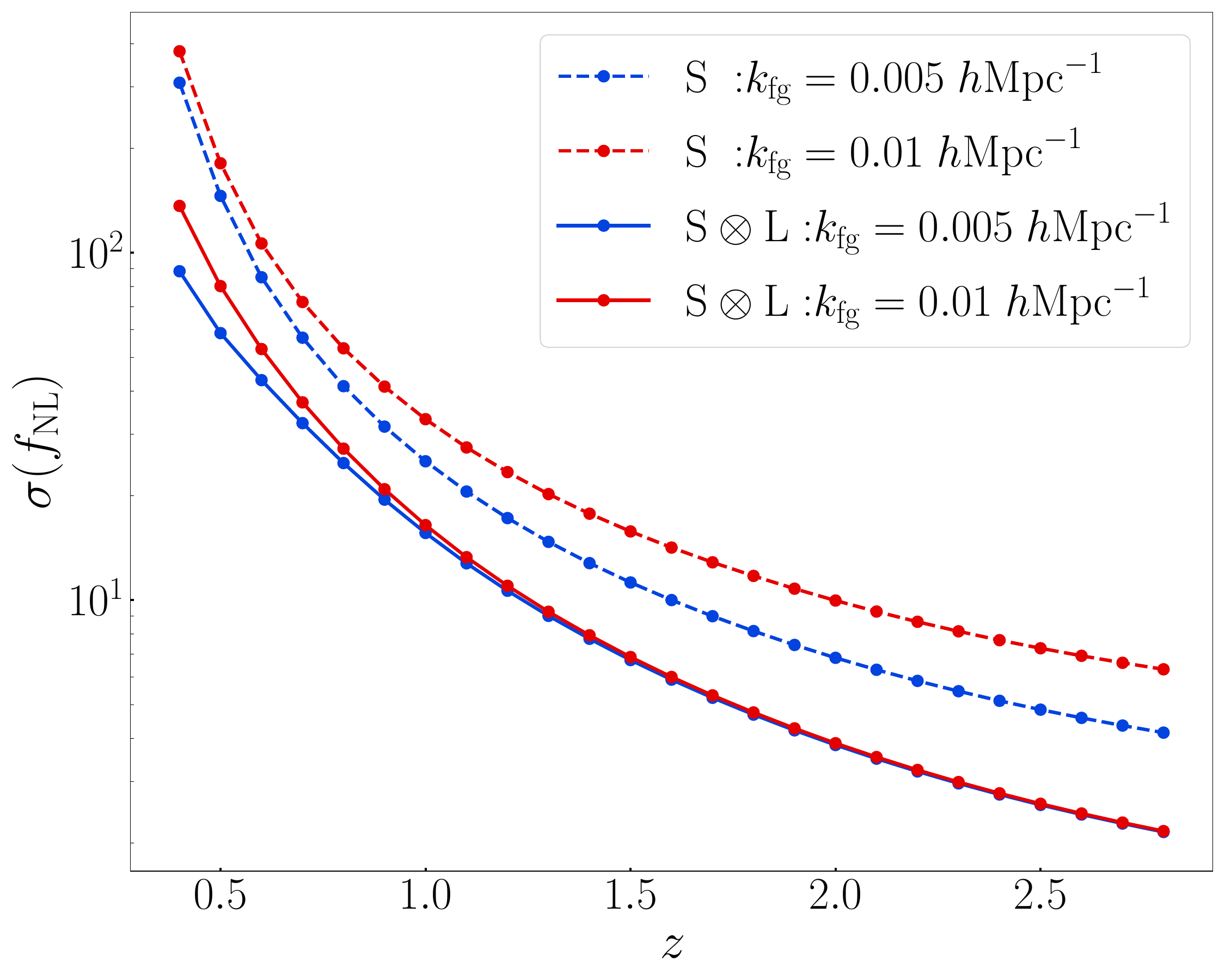}\\
\caption{Cumulative errors on $\fnl$ for the multi-tracer involving single-dish mode HI intensity maps. Results are shown for the two cases of foreground loss, $k_{\rm{fg}}=0.005, 0.01\,h$/Mpc.}
\label{fig:8}
\end{figure}

\begin{figure}[!htbp]
\centering
\includegraphics[width=7.5cm]{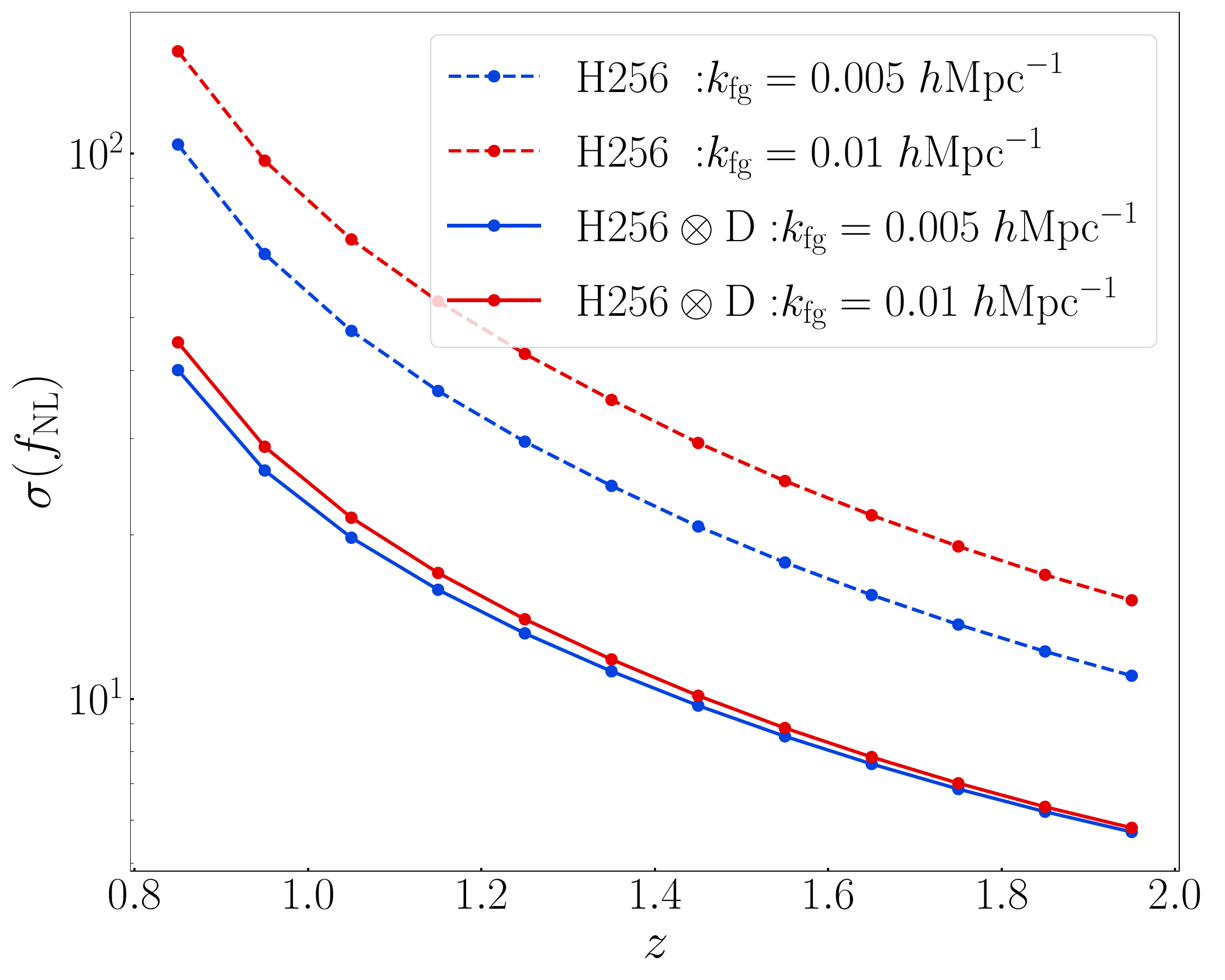}
\includegraphics[width=7.5cm]{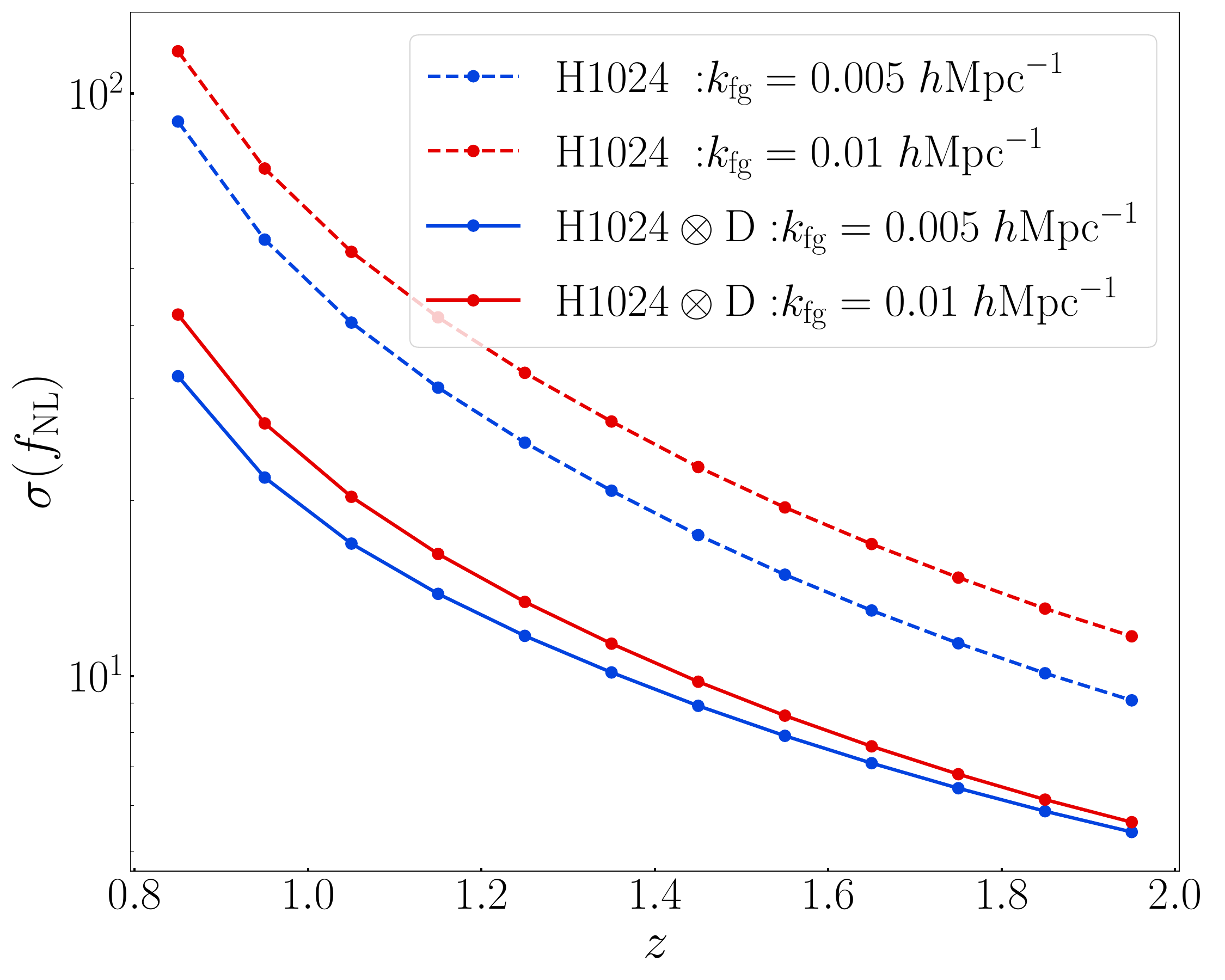}\\
\includegraphics[width=7.5cm]{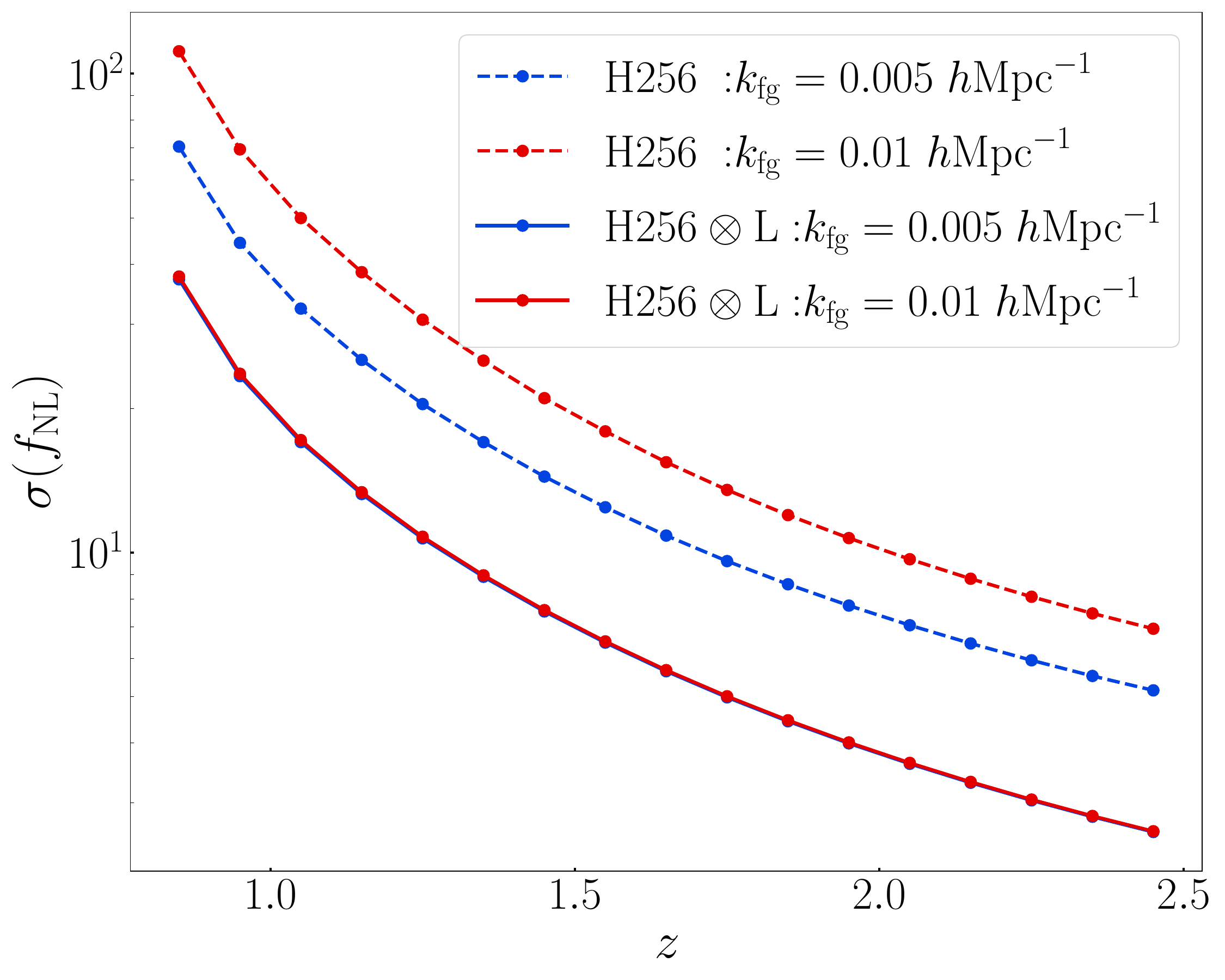}
\includegraphics[width=7.5cm]{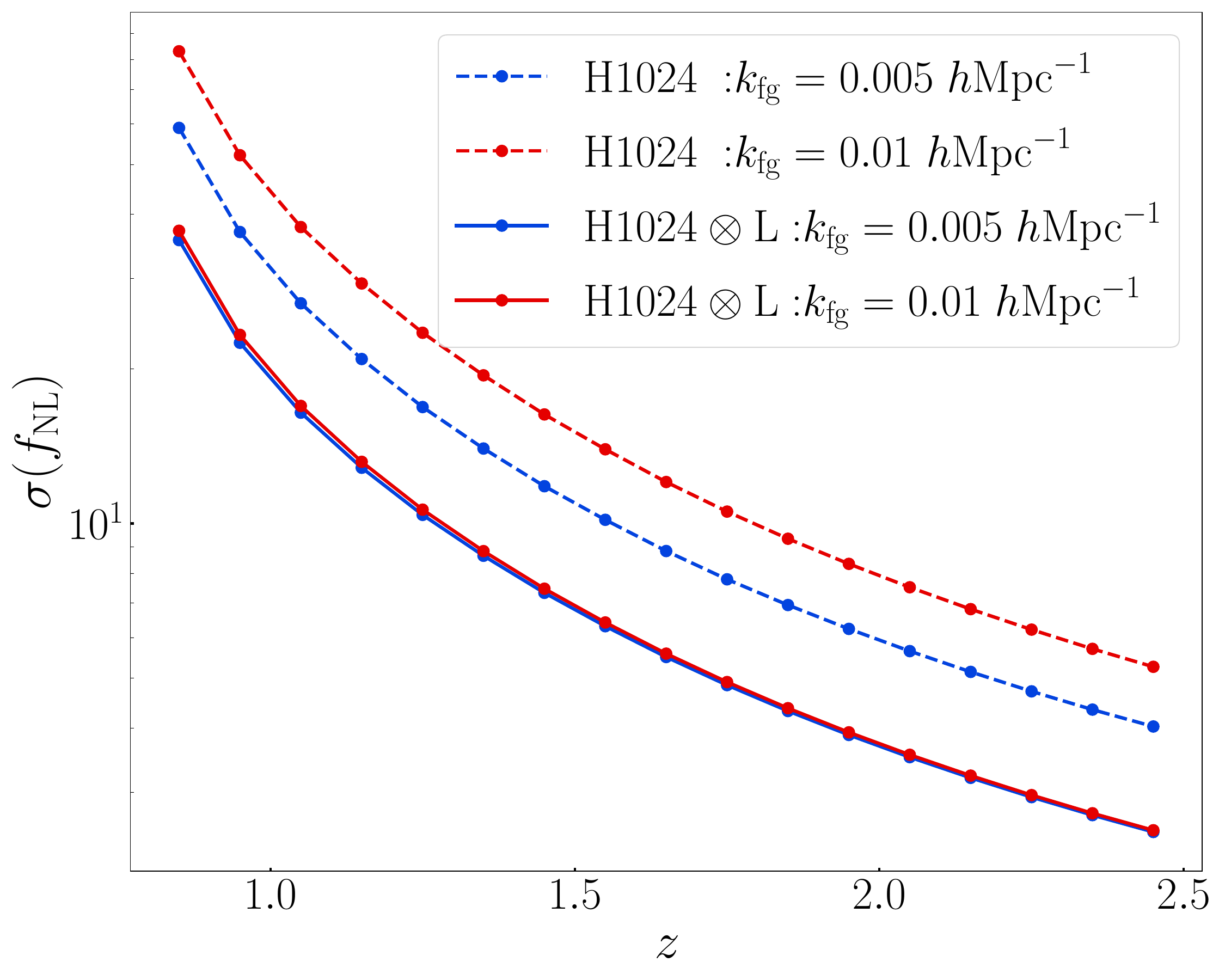}\\
\includegraphics[width=7.5cm]{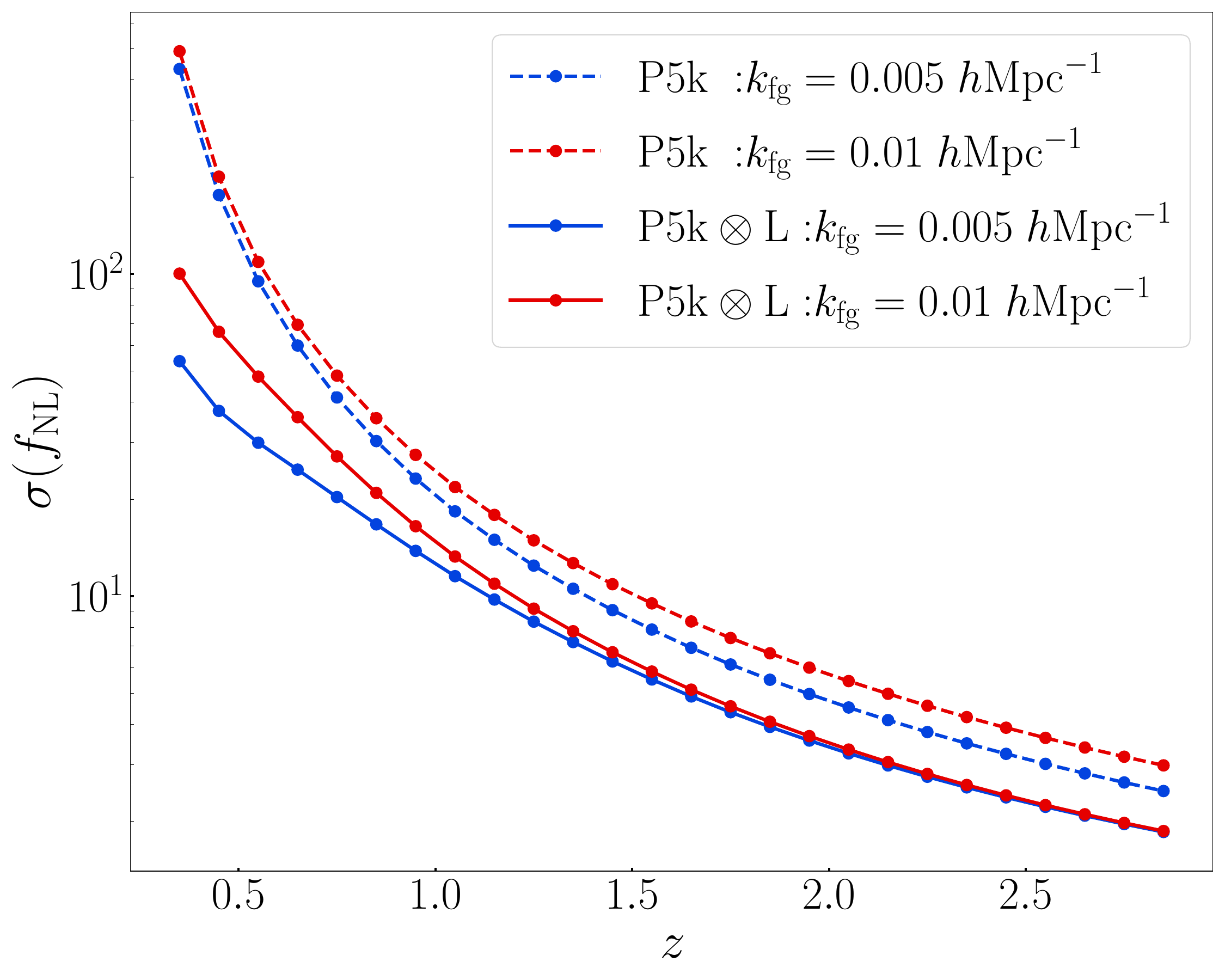}
\includegraphics[width=7.5cm]{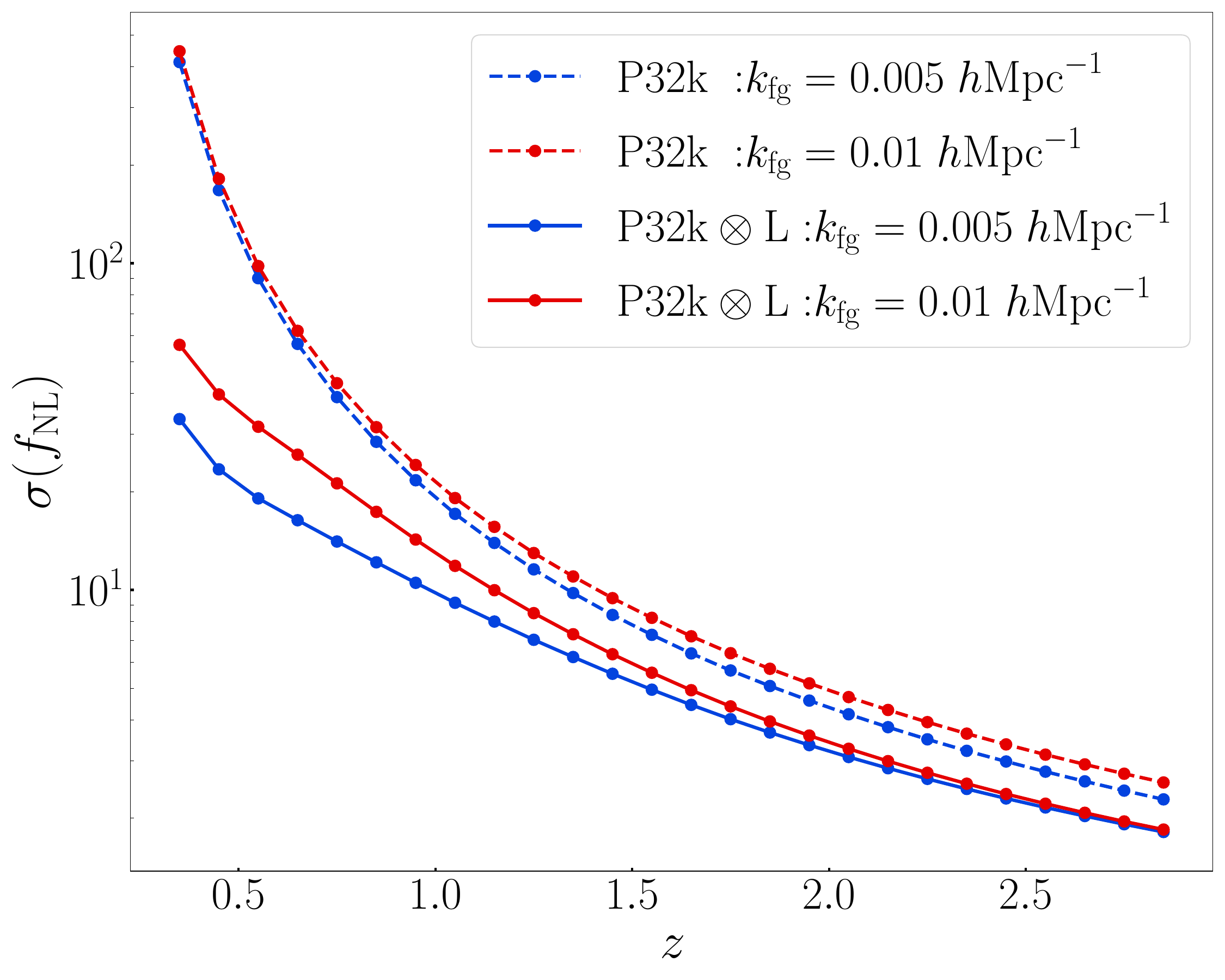}\\
\caption{As in \autoref{fig:8} for the case of HI intensity maps in interferometer mode.
} \label{fig:9}
\end{figure}

The $\sfnl$ results in \autoref{tab3} and \autoref{tab4} are cumulative --  {and they use the optimistic value of the radial foreground avoidance parameter, $k_{\rm fg}=0.005\,h$/Mpc}. The per-bin constraints are shown in
\autoref{fig:8} and \autoref{fig:9}, for the optimistic and less optimistic values  $k_{\rm fg}=0.005, 0.01\,h$/Mpc.


In \autoref{tab5}, we present the results of the multi-tracer analysis for the particular example of H256 $\otimes$ L, illustrating how effectively the multi-tracer approach can mitigate the effects of foreground removal when estimated at a fiducial value of $k_{\rm{fg}} = 0.005\,h$/Mpc or a less optimistic value of $k_{\rm{fg}} = 0.01\,h$/Mpc. We  vary the overlapping sky area  and fix the overlapping redshift range to the fiducial $0.8\leq z\leq 2.5$ (top of the Table). Then we vary the overlapping redshift range  at fixed overlapping sky area  $10, 000\; \rm{deg}^{2}$ (bottom of the Table). The results indicate that as the sky area coverage or redshift range decreases, the impact of the foreground effect becomes more noticeable.

\begin{figure}[!htbp]
\centering
\includegraphics[width=7.5cm]{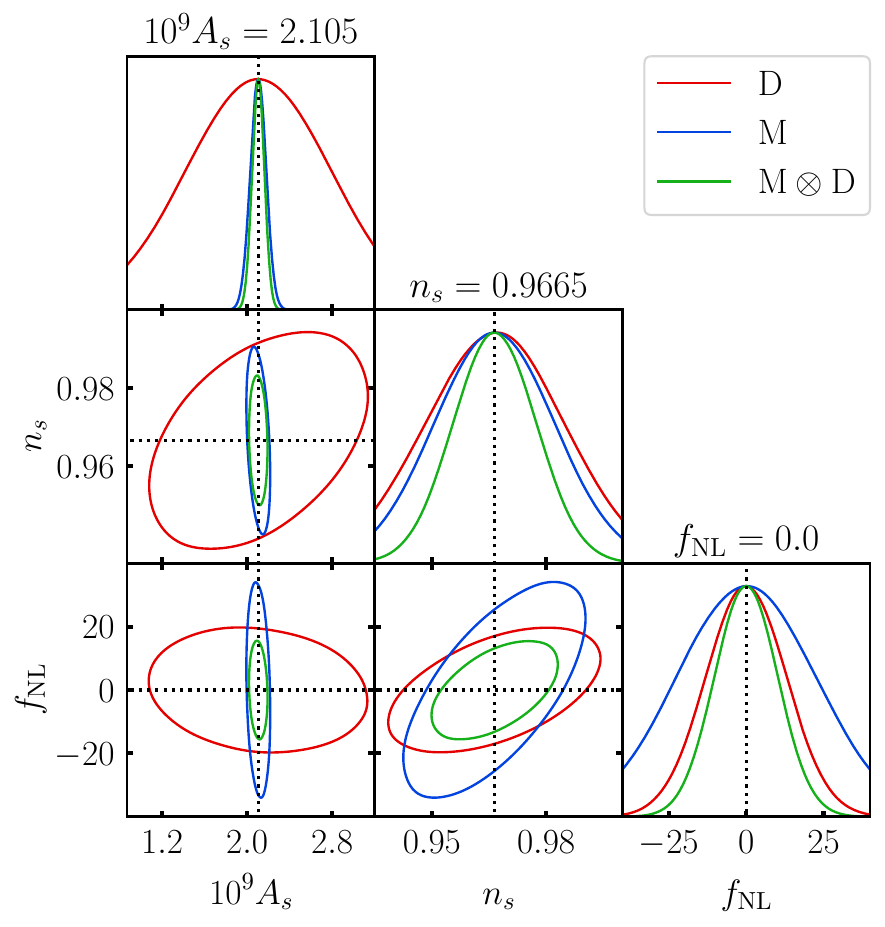}\\
\includegraphics[width=7.5cm]{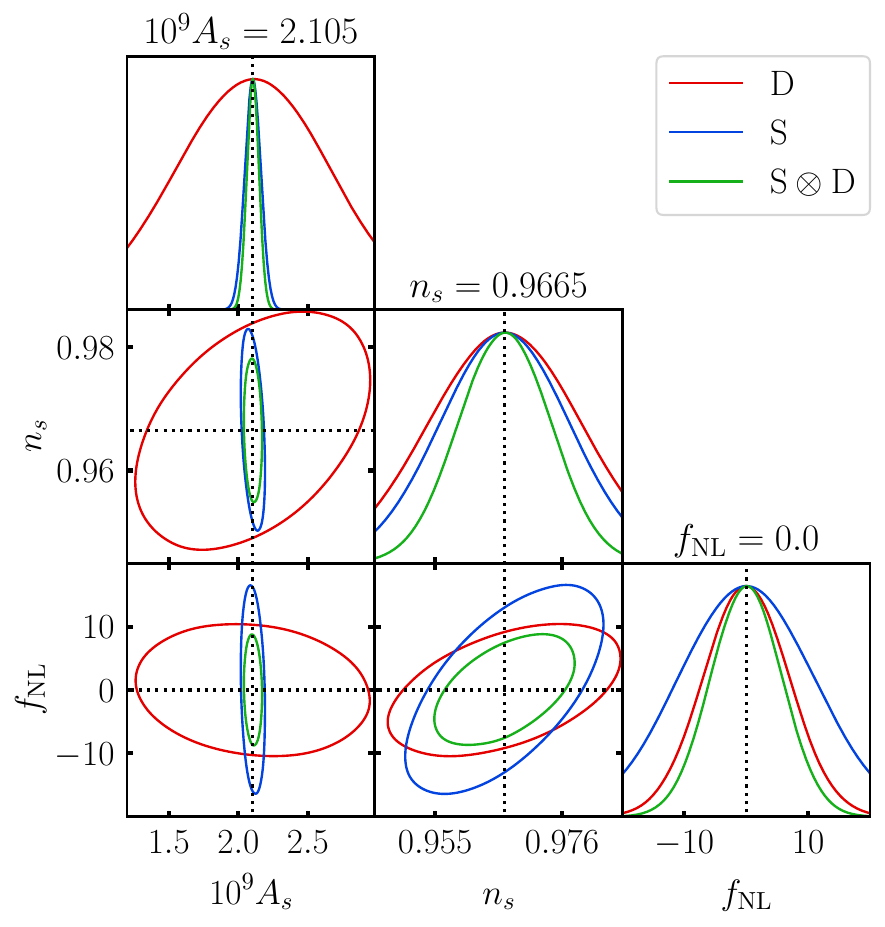}
\includegraphics[width=7.5cm]{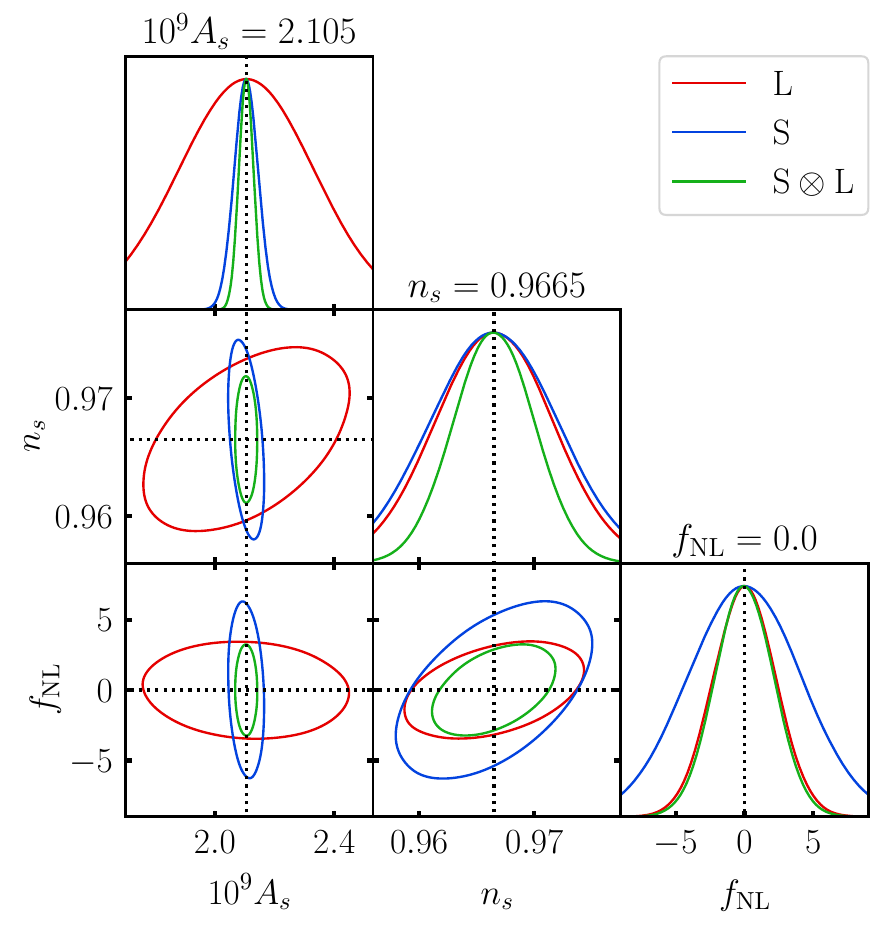}\\
\caption{$1\sigma$ ($68\% $CL) contours for the cosmological parameters, marginalised over the nuisance parameters, in the case of HI intensity mapping in single-dish mode combined with photometric redshift surveys, and with   $k_{\rm{fg}}=0.005;h;\rm{Mpc^{-1}}$.
} 
\label{fig:10}
\end{figure}

\begin{figure}[!htbp]
\centering
\includegraphics[width=7.5cm]{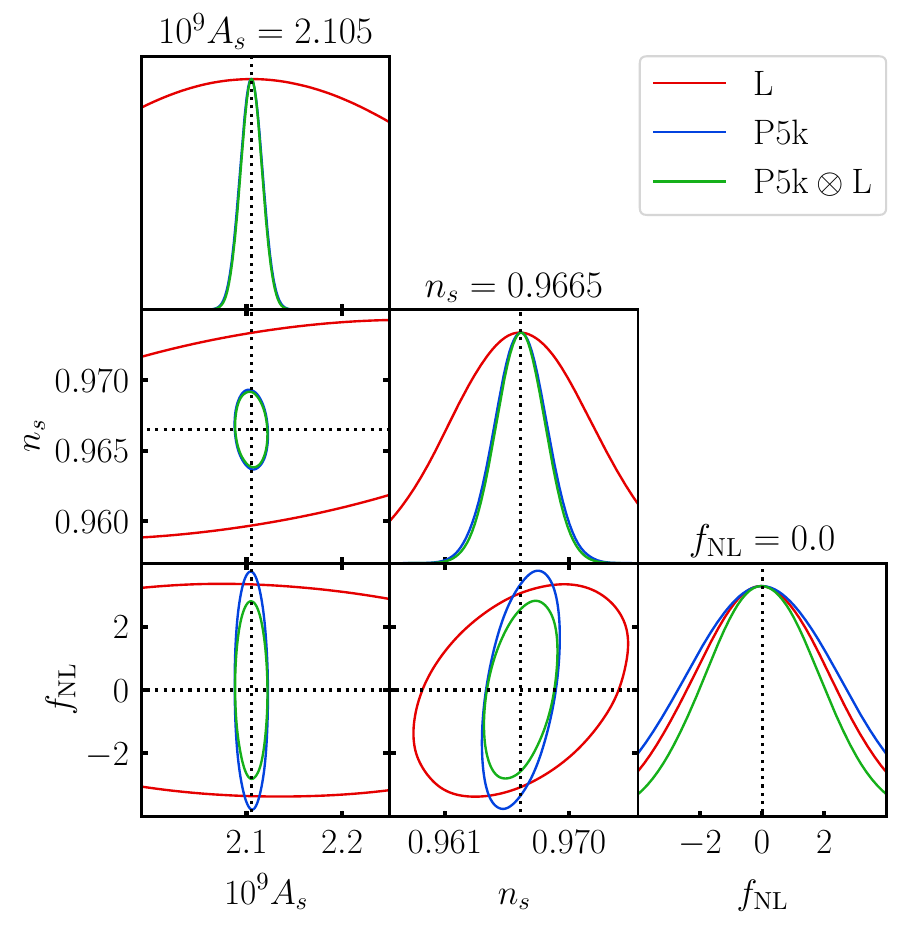}
\includegraphics[width=7.5cm]{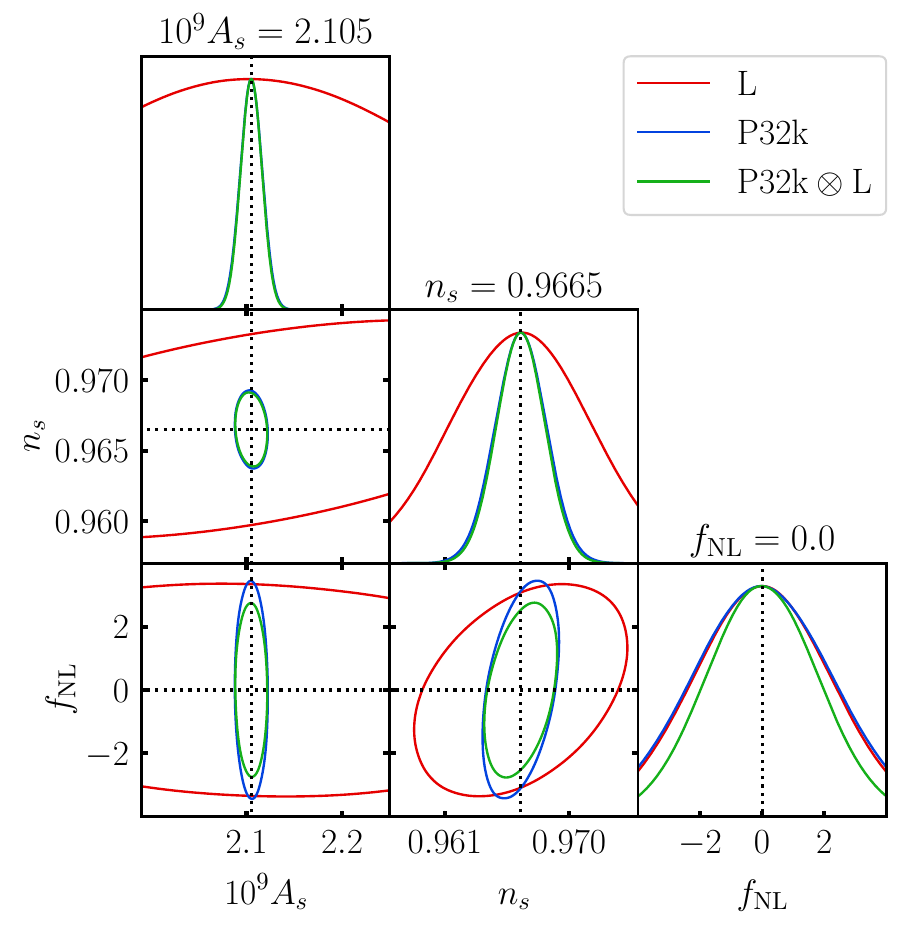}\\
\includegraphics[width=7.5cm]{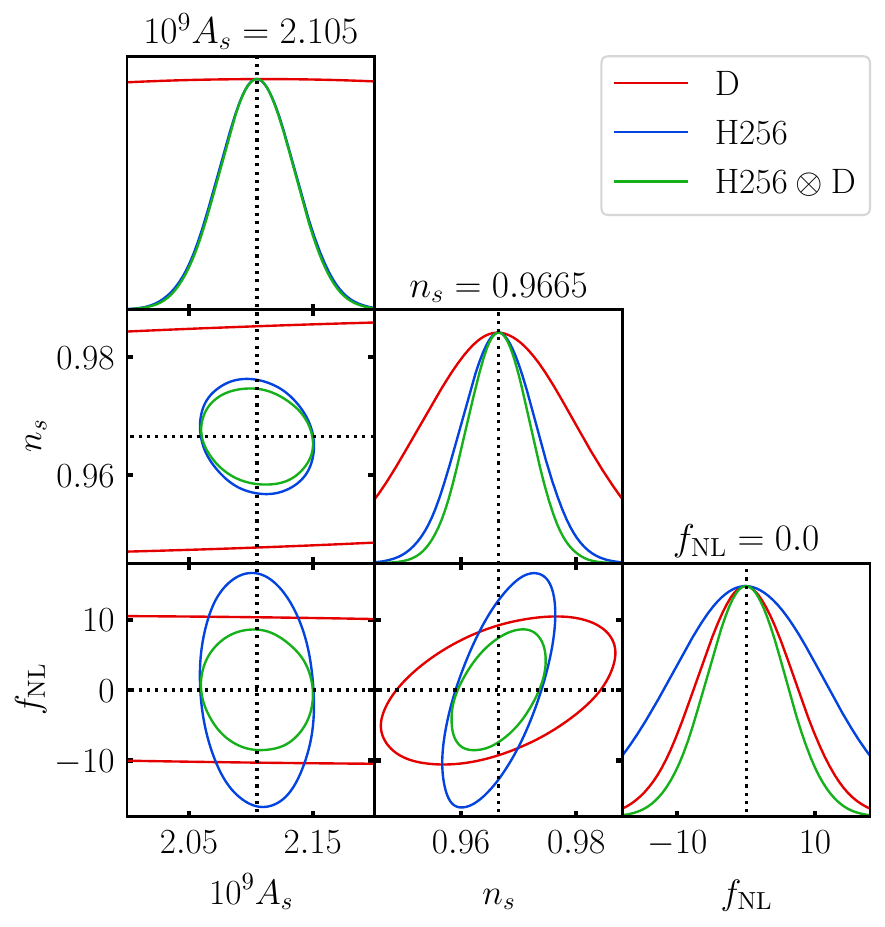}
\includegraphics[width=7.5cm]{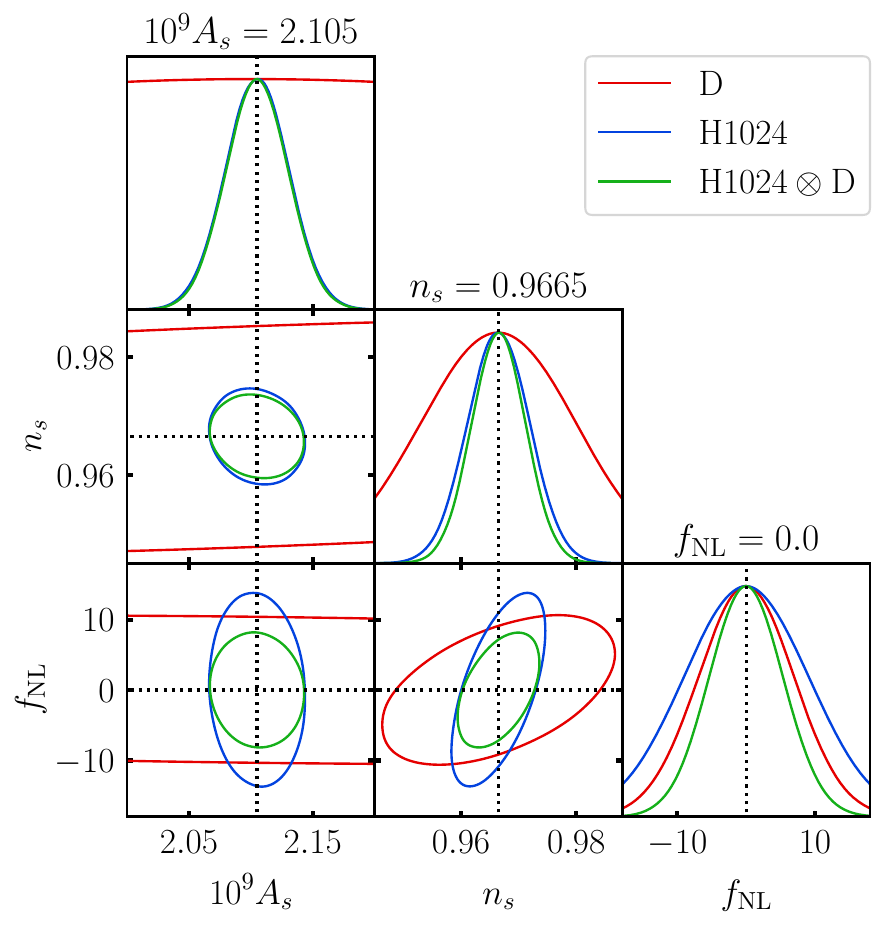}\\
\includegraphics[width=7.5cm]{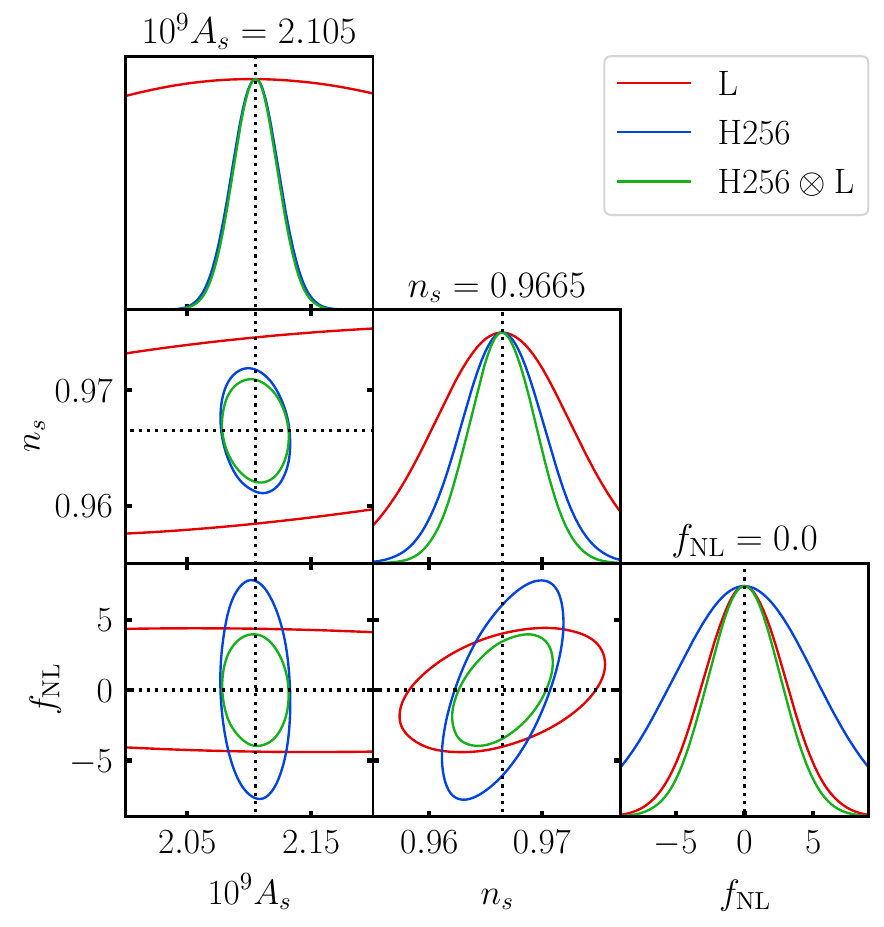}
\includegraphics[width=7.5cm]{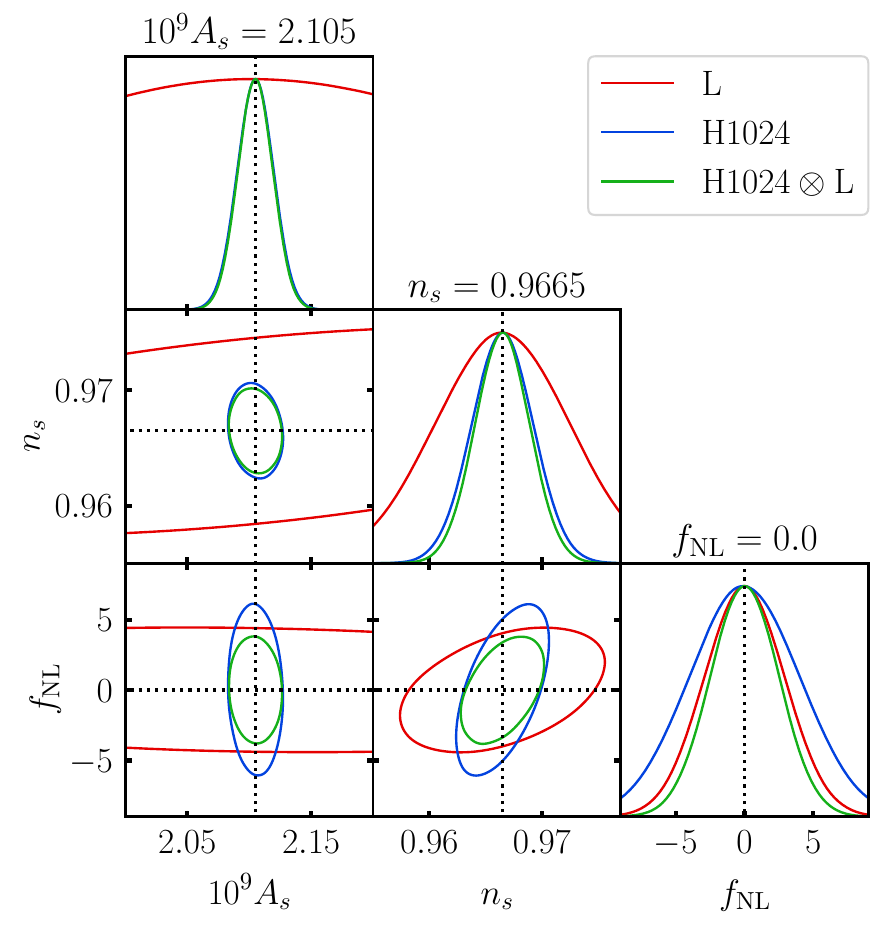}\\
\caption{As in \autoref{fig:10} for the case of HI intensity maps in interferometer mode.
} \label{fig:11}
\end{figure}

\autoref{fig:10} and \autoref{fig:11} present the $1\sigma$ ($68\%$) error contours for $\fnl$, $n_{s}$ and $A_{s}$, marginalising over $b_{A0}$. It is evident that HI intensity mapping offers stronger constraints on $n_s$ and $A_s$ than photometric surveys, although it provides weaker constraints on $\fnl$.

\begin{table}[!htbp]  
\caption{Marginalized $68\%$ (CL) errors on $\fnl$ are analyzed against the estimated surveyed area and redshift ranges for the multi-tracer combination of H256 (HIRAX-like) and L (LSST-like) when $k_{\rm{fg}}=0.005\;h{\rm Mpc}^{-1}$ and $k_{\rm{fg}}=0.01~\;h{\rm Mpc}^{-1}$.} 
\centering 
\label{tab5} 
\vspace*{0.4cm}

\begin{tabular}{c}
  {H256 $\otimes$ L} \\
\begin{tabular}{c r r r}  
\hline\hline   \\ [-0.8ex]
   
 $\Omega_{\rm{sky}}\;[\rm{deg}^{2}]$ & $5,000$ & $10,000$ &  $15,000$
\\ [0.8ex]
\hline  
$\sfnl\;\; [k_{\rm{fg}}=0.005\;h{\rm Mpc}^{-1}]$   & 4.13 & 2.61 & 1.99 \\[1.0ex]
   $\sfnl\;\; [k_{\rm{fg}}=0.01~\;h{\rm Mpc}^{-1}]$   & 4.15 & 2.62 & 2.00   \\[2.0ex]\hline\hline

 $z$ & $0.8-1.5$ & $0.8-2.0$ &  $0.8-2.5$
\\ [0.8ex]
\hline  
$\sfnl\;\; [k_{\rm{fg}}=0.005\;h{\rm Mpc}^{-1}]$   & 7.53 & 3.99 & 2.61 \\[1.0ex]
   $\sfnl\;\; [k_{\rm{fg}}=0.01~\;h{\rm Mpc}^{-1}]$   & 7.58 & 4.01 & 2.62   \\[2.0ex]\hline\hline

\end{tabular} 
 
\end{tabular}  
\end{table} 
\clearpage

\section{Conclusions}\label{Conclusion}

Our analysis shows that combining HI intensity maps and photometric samples allows us to leverage the strengths of both, suppressing the cosmic variance on very large scales, which is an obstacle to precision constraints on local PNG. We emphasise that our goal was not to produce realistic forecasts and develop methods for optimising future measurements.  In particular, we were not aiming to find a multi-tracer combination that could achieve $\sigma(\fnl)<1$.

Instead, we aimed to show the qualitative level of improvement in constraints that could be expected from multi-tracing upcoming surveys that combine surveys with high densities --  photometric and HI intensity mapping surveys. With the exception of the MeerKAT-DES combination \cite{Fonseca:2016xvi} and the SKA-LSST combination \citep{Alonso:2015sfa} (which are consistent with our results),  these multi-tracer combinations have not been previously considered, as far as we are aware.

We find that the relative improvement over the best single-tracer constraints (photometric surveys), are as follows. 

\begin{itemize}
    \item DES-like or LSST-like  $\otimes$ SKA-like versus DES-like or LSST-like alone: $16\%$ and $6\%$, respectively.
   \item DES-like or LSST-like $\otimes$ HIRAX-like versus DES-like or LSST-like alone: 18\% [H256], 23\% [H1024] and 11\% [H256], 14\% [H1024], respectively. 
   
    \item LSST-like $\otimes$ PUMA-like versus LSST-like alone: 17\% [P5k], 18\% [P32k]. 
\end{itemize}

The single-tracer errors in our findings are generally consistent with earlier results, such as those from \citep{Kopana:2023uew} for HIRAX-like and PUMA-like surveys, and \cite{Alonso:2017dgh,Karagiannis:2019jjx,Jolicoeur:2023tcu} for SKA Band 1, or \cite{Alonso:2017dgh,Karagiannis:2020dpq} for SKA Band 1 and the UHF Band of MeerKAT. 

We used a simple Fisher forecast, which produces over-optimistic precision. On the other hand, we discarded the non-overlap information from the multi-tracer pairs, which leads to an under-estimate of precision.

We also investigated the significance of the foreground filtering scale in the multi-tracer analysis. We find that the stronger loss of radial modes in HI intensity maps (e.g., at $k_{\rm fg} = 0.01\,h/\text{Mpc}$) has less impact in the multi-tracer case. For instance, when combining LSST-like and HIRAX-like surveys (red and blue solid lines in \autoref{fig:9}), there is little difference between $k_{\rm fg} = 0.01\,h/\text{Mpc}$ and $k_{\rm fg} = 0.005\,h$/Mpc, because the overlapping sky area and redshift range provide many of the ultra-large scales where the $\fnl$ signal is strongest. By contrast, the MeerKAT-like combination with DES-like (\autoref{fig:8}) is significantly affected by the loss of long modes, due to the smaller overlap in sky area.

For future work, we plan to improve theoretical accuracy by including wide-angle effects (see e.g. \citep{Noorikuhani:2022bwc,Paul:2022xfx,Jolicoeur:2024oij}).

\acknowledgments 
MK and RM are supported by the South African Radio Astronomy Observatory and the National Research Foundation (grant no. 75415). SJ is supported by the Stellenbosch University Astrophysics Research Group fund.

\clearpage

\bibliographystyle{JHEP}
\bibliography{reference_library}

\end{document}